\documentclass[sigchi]{acmart}

\usepackage{tabularx}

\AtBeginDocument{%
  \providecommand\BibTeX{{%
    \normalfont B\kern-0.5em{\scshape i\kern-0.25em b}\kern-0.8em\TeX}}}

\copyrightyear{2024}
\acmYear{2024}
\setcopyright{acmlicensed}\acmConference[CHI '24]{Proceedings of the CHI Conference on Human Factors in Computing Systems}{May 11--16, 2024}{Honolulu, HI, USA}
\acmBooktitle{Proceedings of the CHI Conference on Human Factors in Computing Systems (CHI '24), May 11--16, 2024, Honolulu, HI, USA}
\acmDOI{10.1145/3613904.3642773}
\acmISBN{979-8-4007-0330-0/24/05}

\begin{document}

\title[CodeAid: Design and Semester-Long Deployment of an LLM-based Programming Assistant]{CodeAid: Evaluating a Classroom Deployment of an LLM-based Programming Assistant that Balances Student and Educator Needs}

\author{Majeed Kazemitabaar}
\orcid{0000-0001-6118-7938}
\affiliation{%
  \institution{Department of Computer Science, University of Toronto}
  \city{Toronto}
  \state{Ontario}
  \country{Canada}
}
\email{majeed@dgp.toronto.edu}

\author{Runlong Ye}
\orcid{0000-0003-1069-2795}
\affiliation{%
  \institution{Department of Computer Science, University of Toronto}
  \city{Toronto}
  \state{Ontario}
  \country{Canada}
}
\email{harryye@dgp.toronto.edu}

\author{Xiaoning Wang}
\orcid{0000-0003-1069-2795}
\affiliation{%
  \institution{Department of Computer Science, University of Toronto}
  \city{Toronto}
  \state{Ontario}
  \country{Canada}
}
\email{xiaoningwang@dgp.toronto.edu}

\author{Austin Z. Henley}
\orcid{0000-0003-1069-2795}
\affiliation{%
  \institution{Microsoft Research}
  \city{Redmond}
  \state{Washington}
  \country{USA}
}
\email{austinhenley@microsoft.com}

\author{Paul Denny}
\orcid{0000-0002-5150-9806}
\affiliation{%
  \institution{The University of Auckland}
  \city{Auckland}
  \country{New Zealand}
}
\email{paul@cs.auckland.ac.nz}

\author{Michelle Craig}
\orcid{0000-0003-1069-2795}
\affiliation{%
  \institution{Department of Computer Science, University of Toronto}
  \city{Toronto}
  \state{Ontario}
  \country{Canada}
}
\email{mcraig@cs.toronto.edu}

\author{Tovi Grossman}
\orcid{0000-0002-0494-5373}
\affiliation{%
  \institution{Department of Computer Science, University of Toronto}
  \city{Toronto}
  \state{Ontario}
  \country{Canada}
}
\email{tovi@dgp.toronto.edu}

\renewcommand{\shortauthors}{Kazemitabaar, et al.}

\begin{abstract}
Timely, personalized feedback is essential for students learning programming. LLM-powered tools like ChatGPT offer instant support, but reveal direct answers with code, which may hinder deep conceptual engagement. We developed \textit{CodeAid}, an LLM-powered programming assistant delivering helpful, technically correct responses, without revealing code solutions. CodeAid answers conceptual questions, generates pseudo-code with line-by-line explanations, and annotates student’s incorrect code with fix suggestions. We deployed CodeAid in a programming class of 700 students for a 12-week semester. A thematic analysis of 8,000 usages of CodeAid was performed, further enriched by weekly surveys, and 22 student interviews. We then interviewed eight programming educators to gain further insights. Our findings reveal four design considerations for future educational AI assistants: \textbf{D1)} exploiting AI's unique benefits; \textbf{D2)} simplifying query formulation while promoting cognitive engagement; \textbf{D3)} avoiding direct responses while encouraging motivated learning; and \textbf{D4)} maintaining transparency and control for students to asses and steer AI responses.
\end{abstract}



\maketitle

\section{Introduction}
An increasing number of students are learning to program, not just in traditional computer science and engineering degrees, but across a wide range of subject areas \cite{guzdial2023scaffolding}. Numerous successful initiatives have been developed to broaden participation in computing, for example, by combining computing majors with disciplines in which there has traditionally been greater gender diversity \cite{bradley2022broadening}. However, this surge of interest is putting pressure on resources at many institutions and causing concern amongst administrators and educators \cite{national2018assessing}.  

A particularly challenging aspect involves delivering on-the-spot assistance when students need help. Traditional approaches, such as running scheduled office hours in which students can approach instructors and teaching assistants, are often poorly utilized \cite{smith2017office}. Moreover, in-person support is not equitable as not all students feel comfortable seeking help from an instructor, and students who are bolder may receive help repeatedly while others wait \cite{smith2017my}. There is an urgent need to develop more scalable, equitable and student-friendly solutions for providing support in programming courses.

The recent emergence of large language models (LLMs) may offer one solution. LLM-powered AI tools such as ChatGPT \cite{openai_chatgpt} can act as powerful coding assistants that generate code from natural language descriptions. However, the rapid growth and pervasiveness of LLMs have raised concerns about their use in educational settings~\cite{kasneci2023chatgpt}. This has led to some institutions banning access to tools like ChatGPT \cite{hsu2023should}. In computing education, in particular, concerns have been voiced regarding issues of academic integrity and student over-reliance~\cite{lau2023ban, becker2023programming}. Indeed, research has shown that LLMs can generate direct solutions to almost any problem typical in introductory programming courses \cite{denny2023conversing, finnie-ansley2022robots}. To develop pedagogically effective real-time support solutions for programming courses, it is necessary to implement suitable "guardrails" that restrict the open-ended AI's ability in generating direct solutions even when prompted. This ensures students use the AI constructively \cite{denny2023computing}.

In this paper, we present CodeAid, an LLM-powered coding assistant that is designed to meet the needs of both students and educators: being helpful and technically correct, while not directly revealing code solutions. We used an iterative design approach that involved frequent requirements elicitation and feedback from the course instructor, who taught the course in which CodeAid was deployed. CodeAid allowed students to (i) ask general programming questions, (ii) ask questions about the code they provide, (iii) explain the code they provide, (iv) help fix the code they provide, or (v) help write code.

We deployed CodeAid in a large introductory C and Systems Programming course, spanning a 12-week semester, with about 700 university students. During the first half of the course, we analyzed usage data and provided weekly reports to the course instructor. Halfway through the course, we made several updates to the assistant based on this feedback. Overall, during the semester-long deployment, we collected data from multiple sources: (i) more than 8,000 interactions with CodeAid along with ratings from students for each generated response, (ii) ten weekly surveys about students' usage of CodeAid in comparison with other learning resources, (iii) 22 structured interviews with students discussing CodeAid's features, usability, and helpfulness in learning, and (iv) a final anonymous survey comparing students' usage of CodeAid with other AI coding tools including ChatGPT. We then performed a thematic analysis on 2,100 randomly sampled usages of the system (including the questions that were asked, provided code, and generated responses). Lastly, we presented the results of our deployment to eight University-level programming course instructors from six countries and conducted semi-structured interviews with them to gain further insights into how such AI assistants may be adopted and integrated into new courses in the future.

To effectively understand the implications of AI-powered tools in programming education, this paper is guided by the following research questions:

\begin{itemize}
    \item \textbf{RQ1 - Usage Patterns:} What patterns emerged in student usage of CodeAid, in terms of frequency, choice of features, usage patterns, and the nature of questions posed?
    \item \textbf{RQ2 - CodeAid Responses:} How effective was CodeAid in producing technically correct and helpful responses without directly revealing coding solutions?
    \item \textbf{RQ3 - Student Perspectives:} How did students perceive CodeAid and its comparative advantages over existing tools including ChatGPT?
    \item \textbf{RQ4 - Educator’s Perspectives:} What are the perspectives of educators regarding learner-focused AI assistants like CodeAid in terms of its integration into the curriculum, recommendations for improvement, and effective pedagogy?
\end{itemize}

By synthesizing answers to the above research questions, this paper presents a critical analysis of the broader design space for AI assistants like CodeAid in programming education. We draw on the experience of iteratively developing CodeAid and on the insights from its semester-long deployment to identify four major design considerations for tool design. We highlight the key trade-offs that need to be considered, and present a set of generalizable guidelines for the design of pedagogical LLM-powered coding assistants.
\section{Related Work}
The recent emergence of Large Language Models (LLMs), and their wide array of potential applications, has sparked enormous research interest~\cite{bommasani2021opportunites} and have generated intense debate about the opportunities and challenges they present, especially in domains such as education~\cite{kasneci2023chatgpt, denny2023computing}.  

\subsection{LLMs in Computer Science Education}
As LLMs become more widely used in practice, education researchers are exploring the potential of LLMs to produce educational content, enhance student engagement and customize learning experiences \cite{kasneci2023chatgpt}.  This is especially true in computing education, given that code-generating tools are becoming widely adopted in industry practice.  This has led to ongoing discussions about the need to change how computing is taught~\cite{denny2023computing}.  Instructor opinion on this matter is currently divided. Lau and Guo \cite{lau2023ban} interviewed 20 introductory programming instructors to understand how they plan to adapt their courses. The authors found that in the short term, many educators planned to discourage "AI-assisted cheating" by banning and increasing the weighting of invigilated exam scores,  while others are more willing to embrace AI tools by integrating them into their classes.

Recent work in the computer science education community has started to explore the implications and opportunities of LLMs on computer science learning from different perspectives \cite{becker2023programming}. Most of this recent work has focused on understanding the capabilities of LLMs for completing programming tasks \cite{denny2023conversing} and on generating instructional content \cite{leinonen2023using}. For example, Finnie-Ansley et al. \cite{finnie2023my} showed that Open AI Codex performs better than most students on code writing questions in both CS1 and CS2 exams. Similarly, Savelka et al. compared the capabilities of GPT-3 and GPT-4 on 599 programming exercises from three Python programming courses and found that the GPT models evolved from completely failing the typical programming class’ assessments (the original GPT-3) to passing the courses with no human involvement (GPT-4) \cite{savelka2023thrilled}. In terms of generating learning resources, early work by Sarsa et al. \cite{sarsa2022automatic} analyzed the novelty, plausibility, and readiness of 120 programming exercises generated by OpenAI Codex and proposed the potential of using such models to come up with coding assignments.  In contrast, we explore the use of LLMs to help students complete programming exercises without providing direct code solutions.

\subsection{Question Answering}
Providing accurate and timely answers to student questions is important for effective learning, however, this is a challenge for many computing educators given that class sizes are growing. Moreover, not all students feel equally comfortable approaching an instructor or teaching assistant for help, which can lead to inequity in computing classrooms~\cite{gao2022who}.  The prospect of providing LLM-based support for answering student questions is therefore of great interest to educators~\cite{kumar2023quickta}.  In recent work, Liffiton et al. describe initial work in this direction with their CodeHelp tool which provides assistance to programming students but employs guardrails to avoid directly revealing solutions~\cite{liffiton2023codehelp}.  Students using CodeHelp can enter a free-form question into a text area, along with code and an optional error message. They found that students using CodeHelp over a semester-long programming course (52 students) valued the on-demand availability of the tool, but mostly found it useful for answering specific code-related tasks such as fixing errors. In contrast, CodeAid offers a range of input templates and interactive response formats to cater to diverse student needs. Furthermore, whereas CodeHelp's evaluation centered around student usage and perceptions, CodeAid delves deeper, assessing response quality through thematic analysis and broadening the evaluation scope by involving a larger student cohort (700 students) and gathering insights from course instructors.

\subsection{Explaining Code}
Accurate explanations of code are useful for students learning programming, and can help them improve their reasoning when writing their own code~\cite{murphy2012ability}.  For example, `explain in plain English' questions prompt students to explain their understanding of code at an abstract level~\cite{whalley2006australasian}, and provides both long-term and short-term learning benefits~\cite{vihavainen2015benefits, murphy2012ability}.  Modelling explanations created by experts is an effective way for students to develop this important skill, however, generating high-quality explanations for a large quantity of varying code fragments represents a significant workload for instructors~\cite{macneil2023experiences}.

The generation of code explanations by LLMs is an active area of research. MacNeil et al. \cite{macneil2023experiences} reported student experience with LLM-generated code explanations in a web software development e-book. They showed that most students perceived the code explanations as helpful, but engagement depends on code length, code complexity and explanation types.  Recent work by Leinonen et al. \cite{leinonen2023comparing} showed that LLMs can generate code explanations that are more accurate and easier to understand than those created by students themselves, thus providing a potential scalable solution when compared to peer-generated approaches. CodeAid enhances this by allowing students to ask questions directly from their code to gain clarity on specific concepts, while also offering an interactive feature for line-by-line code explanations.

\subsection{Writing and Debugging Code}
The ability to write code has been a traditionally important learning outcome for novices in introductory programming courses. A common approach for the development of code writing skills has been through the use of frequent programming practice with many small problems~\cite{allen2019many, denny2011codewrite}.  LLMs have shown themselves capable of solving introductory level programming problems with very high accuracy~\cite{finnie-ansley2022robots, reeves2023evaluating}, and thus can provide direct support for code writing when students need help.  The literature on debugging also has a long history, and various tools \cite{head2017writing, hartmann2010would} and activities have been proposed to help novices and students learn debugging techniques~\cite{mccauley2008debugging, lee2014gidget, michaeli2019improving, li2019towards}.  Recent work has shown that LLMs have the potential to be used to assist in many aspects of debugging, including producing more understandable programming error messages \cite{leinonen2023using} and providing high-precision feedback on code for fixing syntax errors \cite{phung2023generating}. CodeAid uses a similar, high-precision method for feedback generation, but also tries to improve the experience by visually annotating the erroneous sections of students' code with suggestions for corrections.

Kazemitabaar et al. developed Coding Steps to explore the use of LLM-based code generators for supporting learners in introductory programming~\cite{kazemitabaar2023studying}.  Coding Steps incorporates a code generator into the user interface of an online programming tool. Students can generate code by providing a natural language prompt to the tool, which is then sent to the OpenAI Codex API, and the response is automatically pasted into the student's code editor.  They studied students using Coding Steps to solve a large set of Python programming tasks. One key finding was that students frequently copied the exercise questions as prompts and then used the AI-generated code without making any alterations to it. This reliance on the code generator is suggestive of the over-dependency problem ~\cite{collins2023policy, brusilovsky2023future, chen2021evaluating}. To address this, CodeAid integrates guardrails to restrict the open-ended AI system from generating direct code solutions even if students ask for them. Additionally, it uses scaffolding techniques like interactive pseudo-code and code annotations to support students in transitioning from understanding concepts to independently writing and debugging their code.

\begin{figure*}
    \centering
    \includegraphics[width=1\textwidth]{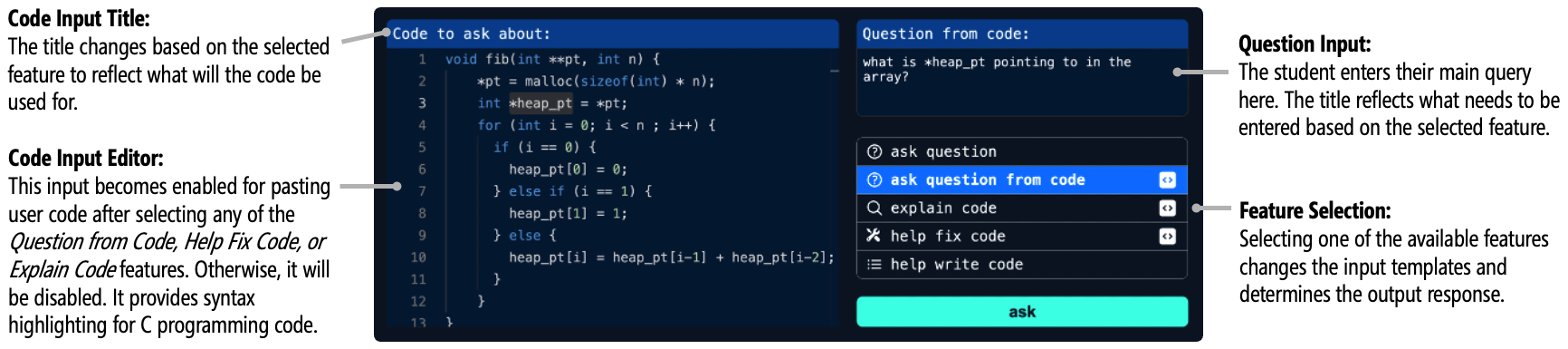}
\Description{This figure presents CodeAid's primary user interface. At its center, the interface displays a code input area on the left, a space for entering questions at the top right, and a set of five radio buttons at the bottom right. These buttons are labelled: "ask question", "ask question from code" (which is currently highlighted), "explain code", "help fix code", and "help write code". Beneath these options, there's an "ask" button. The figure also features four distinct annotations: (1) at the top left, an annotation titled "code input title" indicates the title section of the code input area. It describes that this title adapts based on the chosen feature, providing clarity on how the code will be utilized. (2) Below that, on the left, an annotation named "Code input editor" highlights the space where students can paste their code. This section activates only when "Question from Code", "Help Fix Code", or "Explain Code" are chosen. If not, it remains inactive. Additionally, it offers syntax highlighting specifically for C programming language code. (3) On the top right, the "Question input" annotation points to where students type their main queries. The title of this section alters to guide students on what to input, contingent on the chosen feature. and (4) Lastly, on the bottom right, an annotation called "Feature selection" emphasizes the five radio button choices. Selecting any of these modifies the input prompts and influences the resulting feedback.}
    \caption{The primary input interface of CodeAid. Users select a feature from the bottom right; this choice activates the relevant input fields (code or question). After inputting their query, users press 'ask' and wait for the LLM to respond.}
\label{fig:codeaid_main_input}
\end{figure*}

\section{Initial System Design and Architecture}
CodeAid is an LLM-based programming assistant which aims to assist with programming assignments and reinforce concepts, similar to a teaching assistant, as outlined by Mirza et al \cite{mirza2019undergraduate}.
CodeAid was designed based on prior literature, OpenAI API capabilities, instructor consultations, and pilot studies. The platform has five main features: \textit{Help Write Code} and \textit{Help Fix Code} for hands-on coding support, \textit{General Question} and \textit{Explain Code} for conceptual understanding, and \textit{Question from Code} as a versatile assistance covering both areas. Each feature was carefully designed to produce helpful responses while not directly generating code solutions. In this section, we focus on CodeAid's initial design and architecture. The system went through a major update based on midterm feedback during its deployment which is described later in Section~\ref{section5_system_iteration}.

The interface consists of an \textit{input} to pose questions and an \textit{output} to display past responses. The input section (Figure \ref{fig:codeaid_main_input}) features an input text box, a code input with syntax highlighting, and a radio button group to select from one of the features. A student first selects a feature, enters input (like a function-related question), and hits submit. CodeAid then displays the response using UI elements specific to the selected feature. A section on top of the main input area provided pop-up videos to explain each feature. We also displayed a disclaimer to make sure students understand that the responses are generated by an AI language model and it might generate responses with excessive confidence or be incorrect.

\subsection{Primary CodeAid Assistance Features}
To control the output produced by the LLM and prevent displaying any code solutions to students, we employed few-shot learning as described in \cite{brown2020language}. We provided input/output example pairs to define the overarching format of the LLM's output. This enabled us to confine or restrict code generation, and to add interactive components to the response. This section introduces the design and behaviour of each of the main assistance features.

\subsubsection{General Question}
\raisebox{-2.5pt}{\includegraphics[scale=0.7]{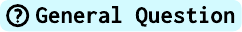}} The most basic feature in our system was generating answers to programming questions, specifically conceptual C programming questions. When a user selects \textit{General Question}, the code editor will become disabled indicating that the AI will only consider the question input. For generating the response, we used few-shot learning to generate short answers with informative explanations. See Figure \ref{fig:v1_prompt_design}a for more details about the prompt design for this feature. The response, displayed in Figure \ref{fig:v1_features}a, was limited to natural language, although sometimes included inline code (such as a function prototype), but no multi-line code.

\subsubsection{Inline Code Exploration}
\begin{figure*}
    \centering
    \includegraphics[width=1\textwidth]{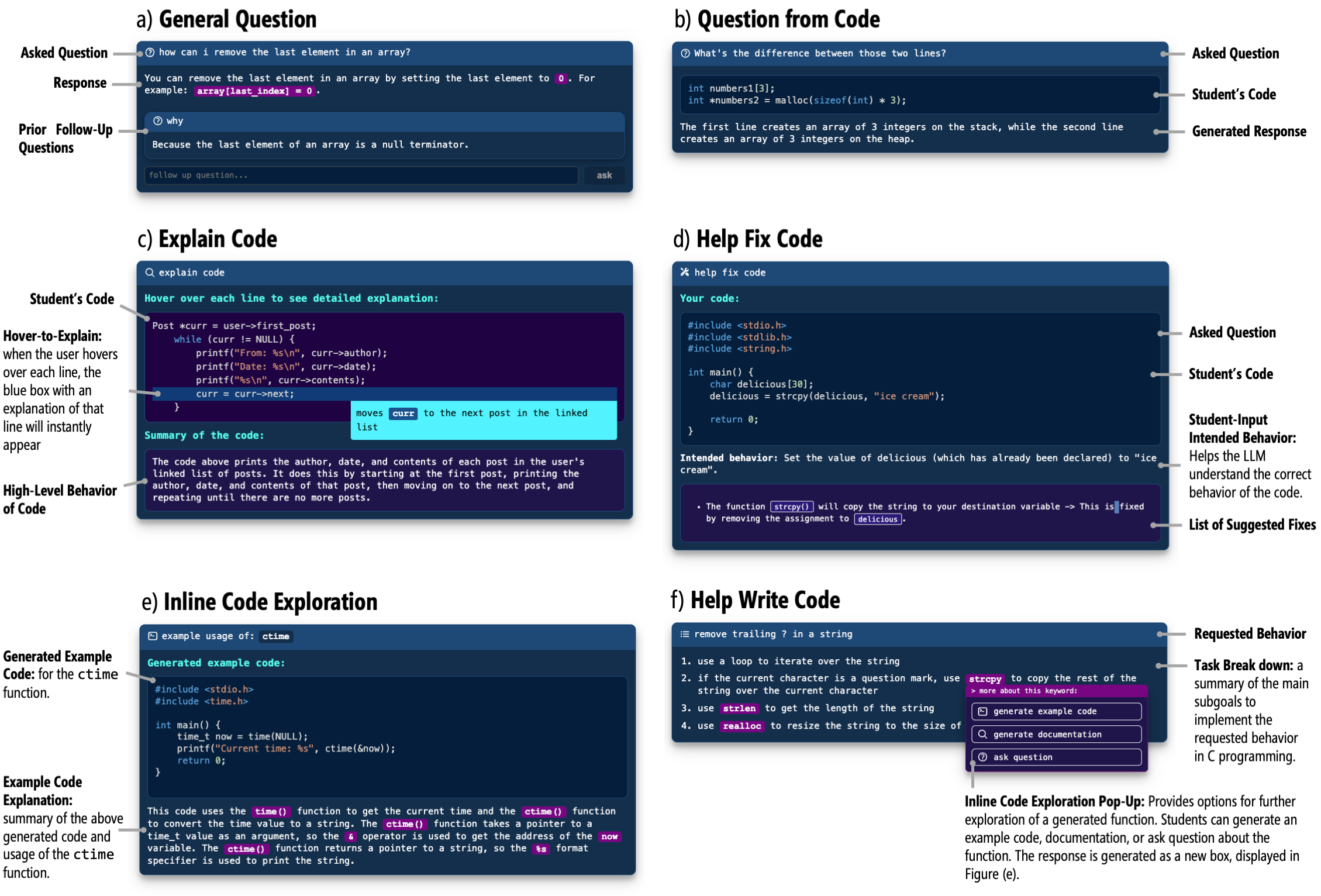}
    \Description{The figure displays the initial versions of responses developed prior to the midterm update, divided into six sub-figures labelled (a) through (f). Sub-figure (a) shows the "General Question" feature, displaying a student's query, CodeAid's response in natural language with some inline code, and follow-up questions. Sub-figure (b) demonstrates the "Question from Code" feature, including the student's query, their code, and CodeAid's response in natural language. Sub-figure (c) displays the "Explain Code" feature, where the student provides a code segment and hovers over a line that displays a blue box with a corresponding explanation for that specific line. Sub-figure (d) highlights the "Help fix Code" feature, presenting the student's code to be fixed, the intended behavior of the code, followed by a bullet-point list of fix suggestions from CodeAid. Sub-figure (e) displays the response from "Inline Code Exploration" usages, which displays a sample use case in C programming code for a specific keyword (here, the 'ctime' function). Finally, sub-figure (f) illustrates the "Help Write Code" feature. Here, the student indicates the desired behaviour, and CodeAid produces a task breakdown, including the main subgoals to implement the requested behaviour in C programming. The interface also displays hovering over a C programming keyword within the natural language response that has opened an Inline Code Exploration pop-up with three options for further exploration of the keyword: generate an example code, documentation, or ask a question about the function. The response is generated as a new box, displayed in sub-figure e.}
    \caption{The initial interface for the responses produced by CodeAid's five primary functions, along with the Inline Code Exploration feature.}
    \label{fig:v1_features}
\end{figure*}

\raisebox{-2.5pt}{\includegraphics[scale=0.7]{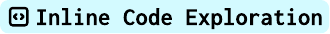}} In each of the primary five features, responses or explanations often contain C programming keywords (such as functions). To provide opportunities for learning, we displayed these keywords in a different style. With this feature, students can hover over a keyword for further exploration: generate sample code about that keyword, generate documentation, or ask a question about the keyword. Invoking any of these three options, will generate an \textit{Inline Code Exploration} response which is displayed in Figure \ref{fig:v1_features}e.

\subsubsection{Question from Code}
\raisebox{-2.5pt}{\includegraphics[scale=0.7]{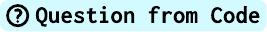}} To simulate an experience similar to StackOverflow's Q\&A forum, we designed the \textit{Question from Code} feature to help students with debugging tasks or conceptual questions in a specific context. The UI for this feature looked similar to the \textit{General Question} feature but with the added ability to provide some code as context (Figure \ref{fig:v1_features}b). Both the code editor and the question input became enabled when this feature was selected.

\subsubsection{Help Fix Code}
\raisebox{-2.5pt}{\includegraphics[scale=0.7]{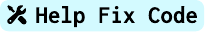}} To help students in their code debugging tasks, students could enter their buggy code in the code editor and the intended behaviour or the problem of the code in the question input (which displayed "Intended Behavior" as its title when this feature was selected). The initial version of the \textit{Help Fix Code} feature (Figure \ref{fig:v1_features}d) performed two tasks in the backend: first, it attempted to generate the correct version of the provided code based on the given description, and second, it tried to explain using bullet points what was changed and why (see Figure \ref{fig:v1_prompt_design}b for more details). The response interface (Figure \ref{fig:v1_features}d) only displayed the bullet points and not the fixed code as a way to not reveal direct code solutions.

\subsubsection{Explain Code}
\raisebox{-2.5pt}{\includegraphics[scale=0.7]{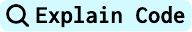}} To help students in use cases such as understanding starter code used in assignments or code taught during lectures, we designed the \textit{Explain Code} feature (Figure \ref{fig:v1_features}c). Upon selecting this feature, the code editor would be enabled and students could paste in code they wanted to be explained. The generated output was an interface that displayed the users' code and enabled them to hover over each line to see the detailed explanation for that line and how the line works in orchestration with the rest of the code. To do this, we used a few-shot learning approach and conditioned the model to produce a simple output structure of generating the same code but with an explanation as a specially formatted comment at the end of each line (see Figure \ref{fig:v1_prompt_design}c to see the structure of the prompt). This enabled us to show the explanations directly to their matching code.

\subsubsection{Help Write Code}
\raisebox{-2.5pt}{\includegraphics[scale=0.7]{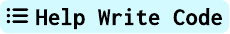}} Prior work has shown LLM-based code generators to provide great starting points for programmers \cite{vaithilingam2022expectation}. Therefore, to assist students struggling with coding tasks, we wanted to provide ways to help them to write code without displaying any code. For that, the \textit{Help Write Code} feature (Figure \ref{fig:v1_features}f) required users to enter the intended behaviour of the program and generated a high-level structure of the code with sub-goals \cite{margulieux2012subgoal} and pseudo-code in natural language. We used few-shot learning to ensure that the generated output included information about C library functions (e.g. for memory allocation or system calls) while not including any code.

\subsection{System Architecture}
CodeAid is written in TypeScript and has a client-server architecture that enables user authentication and storing responses, collecting feedback, and communicating with OpenAI APIs for generating responses. The server is implemented using \textit{Node.js}, specifically: \textit{Express.js} for REST API used in client-server communication, \textit{Mongoose} to interact with a cloud-based instance of MongoDB for storing user data and generated responses, \textit{Passport.js} for user authentication, \textit{Socket.io} for streaming data from OpenAI into the backend (to be stored in the database) and parsed for the client to be displayed in the UI. The client-side code was developed using the \textit{React Framework}.

\begin{figure*}
    \centering
    \includegraphics[width=1\textwidth]{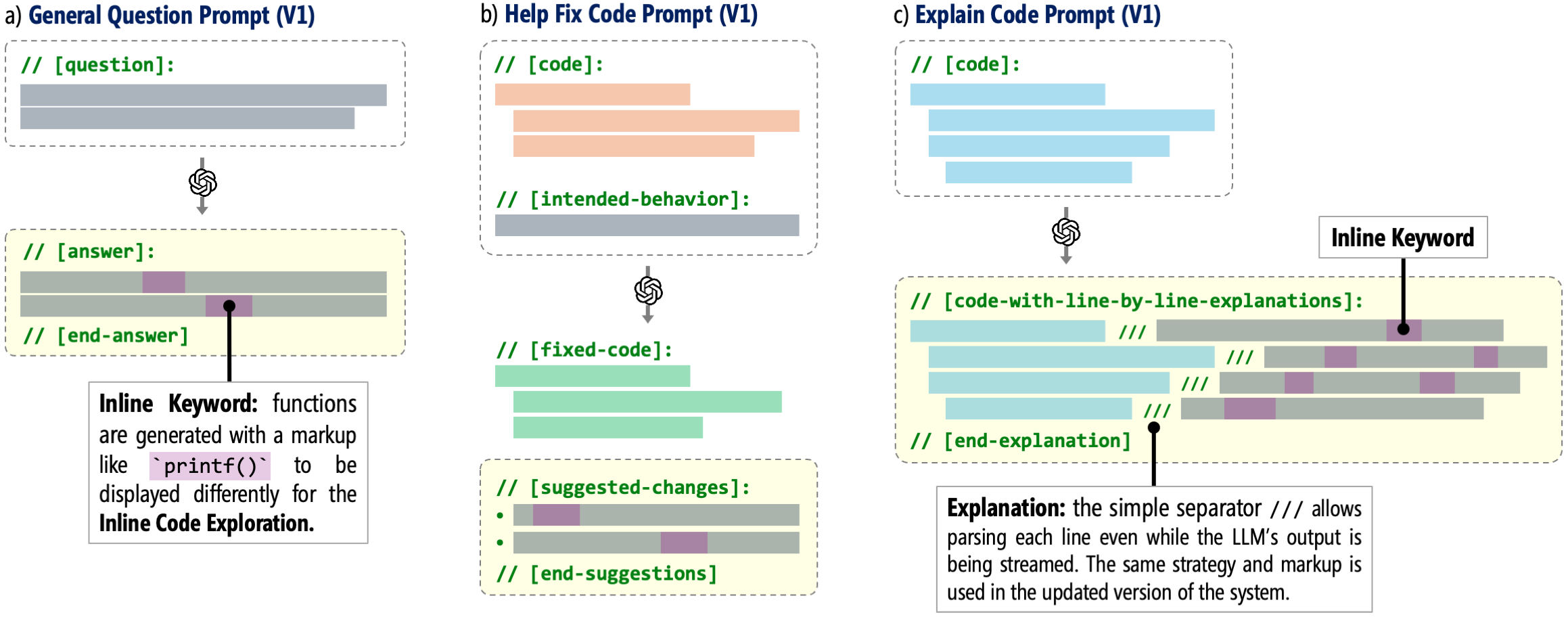}
    \Description{Three sub-figures: (a) \textit{General Question Prompt (V1)}: the prompt starts with '// [question]:' followed by student's query. This is passed to the LLM to receive a block that starts with '// [answer]:' which enables parsing the response. The response can potentially include inline keywords which are enclosed in backticks. Sub-figure (b) is titled \textit{Help Fix Code Prompt (V1)}: the prompt starts with '// [code]:' to denote the beginning of students' buggy code, followed by '// [intended-behavior]' to denote the student's intended behavior for the provided code. This is sent to the LLM to initially display the fixed code, and then to generate the suggested changes (as a list of bullet points). Sub-figure (c) is titled \textit{Explain Code Prompt (V1)}: the input is student's code, and the output is each line of student's code followed by a '///' and a gray bar that indicates the explanation for that line. As mentioned in a callout on the bottom right, "The simple separator '///' allows parsing each line even while the LLM's output is being streamed. The same strategy and markup is used in the updated version of the system."}
    \caption{The structure of LLM prompts used in the initial version of \textit{General Question}, \textit{Help Fix Code}, and \textit{Explain Code}.}
    \label{fig:v1_prompt_design}
\end{figure*}

For user input, we included a textbox and an instance of the \textit{Monaco Editor} that provided syntax highlighting for C programming. These two inputs were selectively enabled or disabled based on the selected feature to indicate which one the user needs to fill in. After users entered the required text for a question and clicked on the generate button, the LLM started streaming output to the backend. The backend simultaneously parsed the response and streamed it to the client. Once the stream finished (e.g. after reaching a specified stop token in the stream), the backend stored the entire response in the database and signalled the client about the finished stream.

The system was instrumented with the Hotjar \cite{hotjar} user behaviour analytics tool that provides heatmaps and anonymous session recordings. We also developed an admin dashboard for the course instructor and the researchers to monitor anonymized students' usage and the associated AI-generated responses. 

\subsection{Prompt Design}
Our prompt engineering consisted of mainly few-shot training, in which at least one input/output example was provided for each prompt. We carefully designed prompts to ensure responses (i) follow structured, and easy-to-parse templates, (ii) are technically correct, and (iii) use a style, tone, and level of technicality that is both helpful and not overwhelming for students. For example, in refining the \textit{Help Fix Code} feature, we moved from generating a simple bullet-pointed suggestions to first generating the correct code, followed by suggested fixes. This significantly improved suggestion quality and accuracy. Similarly, in the updated system, we adopted a pseudo-code style that balances between not overly revealing the code's syntax and not being too close to natural language, which might be too long and overwhelming. Furthermore, the initial system used the \raisebox{-2.5pt}{\includegraphics[scale=0.7]{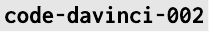}} GPT-3 model, but was updated to the \raisebox{-2.5pt}{\includegraphics[scale=0.7]{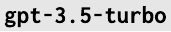}} GPT-3.5 model in the updated system. The structure of the prompts used in the \textit{General Question}, \textit{Help Fix Code}, and \textit{Explain Code} features of the initial version of CodeAid is illustrated in the Figure  \ref{fig:v1_prompt_design}.

\section{Semester-Long Class Deployment}
To gain a comprehensive understanding of students' usage of CodeAid and the AI's generated responses in an authentic learning environment, we deployed CodeAid in a semester-long programming course, with over 700 students, at a large North American university. All students had optional access to CodeAid as an additional resource throughout the semester. Our study, which included the use of an AI tool, interviews, and weekly surveys, was approved by our institute's ethics review board prior to deployment.

\subsection{Course Structure}
The second-year course focused heavily on C programming, and included topics such as shell programming, file processing, processes, signals, system calls, and basic network programming. Prerequisites of the course include a software design course taught in Java and introductory programming using Python. To accommodate the large 700-student class, the course was split into four lectures segments. Students were required to watch specific videos and complete weekly preparation exercises by a set deadline before each lecture. The course included ten lab exercises (worth 1\% each), four programming assignments (A1 to A4, worth 39\% in total), a midterm test (worth 10\%), and a final exam (worth 40\%). To incentivize responding to the weekly surveys about CodeAid, a 0.1\% grade was attributed to responding to each of the ten surveys. This grade was based on completing the surveys, regardless of students' consent to participate in our study. Course policies allowed the use of CodeAid, but explicitly forbade the use of \emph {other} AI tools (such as ChatGPT) to complete any coursework. Students were also asked to cite any external sources that they used for their work. Students in the course had access to a variety of resources beyond CodeAid: recorded lecture videos, lecture notes, an online Q\&A discussion board moderated by the instructors, and weekly office hours with teaching assistants and course instructors.

\begin{table*}[t]
\centering
\caption{The sub-dimensions from our thematic analysis, their associated codes, and inter-rater reliability metrics using Cohen's Kappa and percentage agreement. The detailed codebook is provided in Appendix \ref{appendix_codebook}.}
\begin{tabular}{p{0.3\linewidth}p{0.44\linewidth}p{0.18\linewidth}}
    \toprule
    \textbf{Sub-Dimensions}& \textbf{Codes} & \textbf{Inter-Rater Reliability}\\
    \midrule
    What are students asking from CodeAid?& \textit{Code and conceptual clarification}, \textit{Function-specific queries}, \textit{Code execution probes}, \textit{Buggy code resolution}, \textit{Problem source identification}, \textit{Error message interpretation}, \textit{High-level coding guidance}, and \textit{Direct code requests}& 88\% ($\kappa=.85$) \\
    \midrule
    How much is CodeAid directly revealing the solution?& \textit{Specific code}, \textit{Specific pseudo-code}, \textit{Example high-level code}, \textit{Example high-level pseudo-code}, \textit{Steps to fix syntax issue}, \textit{Steps to fix semantic issue}, \textit{No-code conceptual explanation}& 94\% ($\kappa=.92$) \\
    How technically correct is the response?& \textit{Correct}, \textit{Incorrect}& 87\% ($\kappa=.62$) \\
    How helpful is the response if correct?& \textit{Helpful}, \textit{Not helpful}& 82\% ($\kappa=.61$) \\
    \bottomrule
\end{tabular}
\Description{The table presents a thematic analysis of various sub-dimensions, their respective codes, and inter-rater reliability metrics. The sub-dimensions encompass aspects like the nature of the query, context provision, extent of code solution disclosure, technical correctness, helpfulness if accurate, and the relevance of displayed c-library-functions. For each sub-dimension, specific codes detail the nuances of each theme, such as "code execution probes" under "what is being asked" or "sufficient context" under "how is context provided". The inter-rater reliability for these themes is quantified using percentage agreement and Cohen's Kappa, with values ranging between 82-94\% for agreement and 0.61-0.92 for Cohen's Kappa.}
\label{tab:highlevel_codebook}
\end{table*}

\subsection{Deployment and Participants}
To ensure ethical integrity and avoid any perceived pressure, students' engagement with CodeAid, weekly surveys, and interviews in our study was entirely voluntary and confidential from course instructors. The researchers informed the instructors only at the end of the semester about who had completed the surveys, which contributed 1\% to the course mark, without revealing the participants' consent status. Students' consent was obtained through the first weekly survey, where they selected from three options: consent to share their CodeAid data for research, participate in surveys for the grade without data sharing, or opt out of surveys, foregoing a potential 1\% grade increase. Out of all, 563 (80\%) students consented to participate and share their data for our analysis.

Of the 563 participants, 318 (56\%) reported their gender identity as man, 170 (30\%) woman, 4 (1\%) non-binary, and 71 (13\%) preferred not to say. In terms of English proficiency, 515 (91\%) students agreed or strongly agreed that they are comfortable reading English. In terms of program of study, 418 (74\%) were enrolled in a Computer Science major program, 81 (14\%) were enrolled in a Computer Science minor program, and 64 (12\%) the rest were in other programs. In terms of prior knowledge about C programming, 348 (62\%) students disagreed or strongly disagreed about being competent in C programming before the course, while 57 (10\%) students agreed or strongly agreed. Students were invited to use CodeAid through several email announcements throughout the semester. 

\subsection{Data Sources}
To gather a comprehensive understanding of student and educator experiences and interactions with CodeAid, we employed a multifaceted data collection approach including CodeAid's interaction logs, ten weekly feedback surveys, an anonymous post-course survey (administered after final grade submissions), and semi-structured interviews with 22 randomly sampled students.

\subsubsection{CodeAid Usage and In-situ Feedback}
A primary data source for understanding students' usage patterns (RQ1) and the assistant's response quality (RQ2), was CodeAid’s activity logs. Log data has become an important data source to understand programming experiences \cite{brandt2009two} and coding approaches \cite{finnie2023my, ichinco2015exploring}, particularly when interacting with LLMs \cite{kazemitabaar2023studying, kazemitabaar2023how_novices_use_llm_code}. For each question asked by students, we closely examined its content and CodeAid’s generated responses through a thematic analysis described later in this section. To better understand the usefulness of the AI-generated responses (RQ2), CodeAid prompted students with a mandatory question that asked \textit{"How useful was this response?"}. Students had to respond to a 5-point Likert scale (displayed as a rating stars) and optionally type a reason for their rating before they could use the system again. For each feature, we analyzed the ratings and grouped the reasons into positive and negative feedback to better understand students' perceptions. 

\subsubsection{Weekly Research Surveys}
To understand how students used CodeAid in comparison to traditional educational resources and ChatGPT (RQ3), we conducted a weekly online survey. These types of surveys have aided researchers in monitoring students' perceptions towards AI agents \cite{wang2021towards}. Similarly, we asked students to report their comparative usage of multiple resources including lecture videos, lecture notes, Q\&A discussion boards, office hours, and CodeAid. We also asked questions about why they did or did not use CodeAid, how useful they found it, what they liked or disliked about the tool, and any open-ended feedback about CodeAid during the last week of the course.

\subsubsection{Semi-structured Interviews with Students}
To gain deeper insights about how students used CodeAid (RQ3), we conducted confidential interviews with 22 randomly selected participants, ensuring their privacy from the instructors. Eight interviews were conducted halfway through the course (after students had used CodeAid for A1 and A2), and 14 interviews were done after the course was finished (after their final exams). After obtaining informed consent, our interview questions addressed productivity enhancement, shifts in workflow, verifying responses and reliability, usability concerns, learning moments with CodeAid, and contrasts with alternative resources like StackOverflow, a moderated Q\&A discussion board, and other websites. We concluded the interviews with a short co-design activity, involving students as collaborators in envisioning and shaping future iterations of CodeAid. This approach effectively gathered their unique ideas and suggestions for features tailored to educational settings. Each interview lasted approximately an hour and every participant was compensated with a \$25 gift card.

\subsubsection{Post-Course Anonymous Survey}
Since course policy prohibited the use of code generated from AI tools like ChatGPT to complete any course work, we conducted an anonymous survey to determine whether and why students used ChatGPT. Our goal was to gain a holistic view of students' perceptions of AI assistants in large classes (RQ3). The survey further explored the frequency of their engagement with ChatGPT in comparison to CodeAid. 

\subsection{Thematic Analysis}
To gain deeper insights into how students interacted with CodeAid and the quality of the AI-generated responses, we performed a thematic analysis on the usage logs. From a total of 8132 usages, we initially removed those from students who did not agree to participate in the research, leaving 7003 data points. We then randomly sampled 2100 data points (30\%) and then removed usages that were unrelated to the course (n=71) were excessively long (n=133), or in which CodeAid encountered a technical error (n=82). We were then left with 1750 (25\%) usages, on which we performed the thematic analysis and report our results. 

We created two high-level code dimensions to answer our research questions: (i) to understand usage patterns, choice of features, and the nature of questions posed (RQ1), we focused on the \textbf{\textit{User Query}} (including any provided code or error logs), and (ii) to evaluate the quality of the AI-generated responses (RQ2), we focused on the \textbf{\textit{CodeAid Response}}. This enabled us to focus on relevant data for each of the sub-dimensions in our thematic analysis in the following rounds of analysis \cite{bingham2021deductive}. 

Under each of the two dimensions, we applied an inductive approach where two researchers read through  100 randomly sampled data points together and allowed codes to emerge during the process \cite{bingham2021deductive}. The process involved familiarizing themselves with the data, specifying sub-dimensions, and then creating codes for each sub-dimension. The two researchers then independently coded another 120 randomly sampled data points using the initial codebook. Next, they discussed the results from the initial coding, resolved conflicts, and further refined the codebook. During this step, they presented the codes and representative usages to the course instructor and incorporated their feedback. To enhance the generalizability and reliability of our findings, we streamlined our coding definitions. For instance, responses with minor inaccuracies were labelled as \textit{"incorrect"}. Similarly, if responses that were categorized as \textit{"correct"} failed to adequately address the query, were irrelevant, repetitive, or exceeded the scope of the course material, they were labelled as \textit{"unhelpful"}.

After refining the codebook \ref{tab:highlevel_codebook}, the two researchers independently coded 200 data points and used Cohen's Kappa and percentage agreement \cite{miles1994qualitative, neuendorf2017content} to compute the inter-rater reliability for each of the sub-dimensions. After removing low-quality sub-dimensions, addressing disagreements, and finalizing the codebook, the two researchers independently coded a total of 1430 additional data points selected at random from the remaining untagged data. The full codebook can be found in Appendix \ref{appendix_codebook}.

\section{Midterm Feedback and System Iteration} ~\label{section5_system_iteration}
Midway through the course, following the completion of half the major assignments, we conducted semi-structured interviews with eight randomly selected students and analyzed five of the weekly research surveys. This feedback was used to derive any needed iterations to the design of the tool. 

The feedback highlighted numerous aspects of CodeAid that students appreciated: clear explanations of code or complex topics, assistance in identifying errors, constant availability, direct and personalized engagement compared to online searches and documentation, and the flexibility to ask diverse questions. 
However, our analysis also pinpointed several areas where CodeAid could be improved. Specifically, students felt that the responses they received were often too brief and lacked in-depth information, example usage code, or associated documentation. There were also concerns about incorrect answers or misleading suggestions for fixing their code. Another common frustration was the slow response time of the platform. Students pointed out the importance of the \textit{"Help Fix Code"} feature, but suggested that it should provide specific line numbers where errors were detected. Furthermore, students expressed the need to ask follow-up questions beyond the \textit{"General Question"} feature. Lastly, during the co-design phase of our interviews, students emphasized the need for seamless access to documentation for functions mentioned in CodeAid's responses.

\begin{figure*}
    \centering
    \includegraphics[width=1\textwidth]{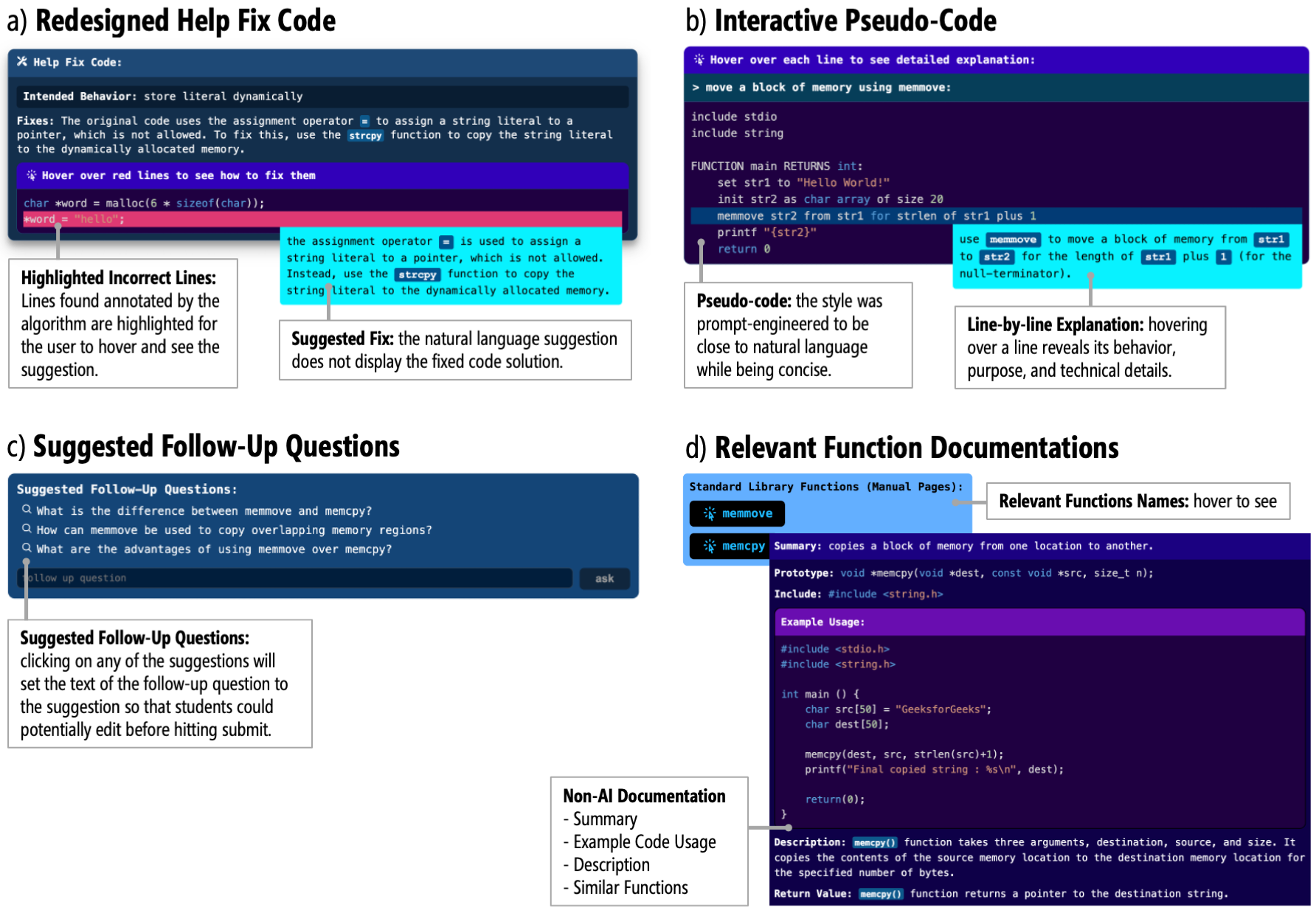}
    \Description{The figure displays the redesigned interface of CodeAid, following the midterm update and contains four primary enhancements: (a) the "Help Fix Code" feature which now visually highlights incorrect lines and allows users to hover over a line and read the natural language suggested fix; (b) an interactive pseudo-code. The format and style of the pseudo-code was prompt-engineered to be close to natural language while also being concise. Students can hover over each line of the pseudo-code to understand more about the behavior, purpose, and technical details of the line; (c) suggested follow-up questions added to the new features with follow-up questions; and (d) Relevant Function Documentations: CodeAid will also suggest a list of relevant function names for each question asked by the student. By hovering over the functions, students can see non Al-generated documentation, containing, summary, example code usage, description, similar functions}
    \caption{The redesigned interface for the responses produced by CodeAid after the midterm update: (a) redesigned \textit{Help Fix Code}, (b) the new interactive pseudo-code with line-by-line explanations, (c) suggested follow-ups, and (d) displaying relevant function definitions.}
    \label{fig:v2_system_features}
\end{figure*}

\subsection{System Updates and Enhancements}
In response to the feedback received during our initial evaluation, and after in-depth discussions with the course instructor, we implemented several updates to CodeAid.

\subsubsection{Pseudo-Code Integration}
As a way to provide more comprehensive responses and to increase engagement, we decided to add pseudo-code generation to most of the features (Figure \ref{fig:v2_system_features}b). We used pseudo-code as a form of scaffolding, offering a simplified and structured outline of a program, which serves as a bridge to actual coding without directly revealing the code itself \cite{lee2015comparing}. To enable this functionality, we updated our LLM prompts for all features to ask the model to generate code. However, instead of showing this code to the user, we sent it to a new LLM function that generated the pseudo-code. For line-by-line explanations, the new LLM function also included an explanation following each line in the form: \raisebox{-2.5pt}{\includegraphics[scale=0.7]{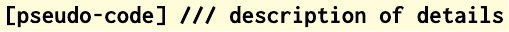}}. This was then parsed and only the pseudo-code was rendered, while the explanation would be displayed whenever the user hovers over that particular line.

\subsubsection{Displaying Function Documentation}
Since the \textit{Inline Code Exploration} was underutilized, we removed it and instead updated all of the features to display static (non-AI generated) documentation for functions that were relevant to the query. Students could hover over the function buttons to see the detailed documentation with usage descriptions and code examples (Figure \ref{fig:v2_system_features}d). We implemented this by asking the LLM to always list all relevant functions to the user's query. To retrieve the documentation for each function, we developed a local key-value database in which the keys were function names and values were the documentation objects scraped from the Standard C Library parsed into a JSON object.

\subsubsection{Stream Generation}
To address concerns over response delays, we incorporated OpenAI's stream generation mechanism for CodeAid's responses to provide immediate feedback. To achieve this, we developed a specific markup for each of the features to enable parsing the partially generated response as it was being streamed. This enabled our system to immediately start displaying responses after users clicked the generate button. For single-line components such as generating a short summary for the provided code, or responding with a single-line answer to an asked question, our markup used the following format: \raisebox{-2.5pt}{\includegraphics[scale=0.7]{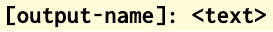}}. For multi-line components such as code parts, we used two tokens, \raisebox{-2.5pt}{\includegraphics[scale=0.7]{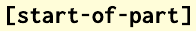}} and \raisebox{-2.5pt}{\includegraphics[scale=0.7]{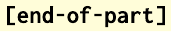}}. Furthermore, in situations where the output of one LLM function was sent to another LLM function in series (e.g. to generate the final pseudo-code from code), our API did not show the generated code, but displayed (and updated while streaming) the number of lines of code that were generated.

\subsubsection{Updated Prompts and OpenAI Model}
We improved the \textit{General Question} and \textit{Question from Code} features to provide more detailed and thorough answers by refining the few-shot prompting examples. To enhance the overall quality, accuracy, and reliability of CodeAid's responses, we upgraded the AI model from \raisebox{-2.5pt}{\includegraphics[scale=0.7]{sections/figures/inline-images/t8-llm-code-davinci.pdf}} to the more advanced \raisebox{-2.5pt}{\includegraphics[scale=0.7]{sections/figures/inline-images/t9-llm-gpt-3.5.pdf}} model. We also updated the prompts to not respond to questions that were not relevant to C programming.

\subsubsection{Redesigned Help Fix Code}\label{redesigned_help_fix_code}
\begin{figure*}
    \centering
    \includegraphics[width=1\textwidth]{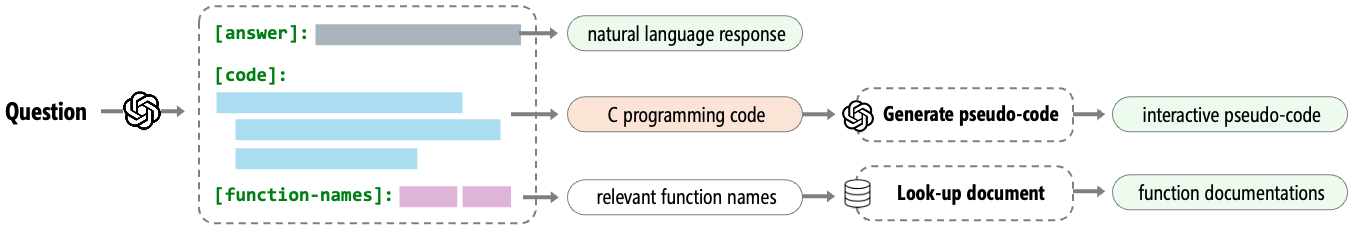}
    \Description{Figure shows revised prompt design and system architecture illustrating the data flow of the updated General Question feature, highlighting the process of generating pseudo-code and relevant function documentation. The Question first gets processed by the LLM, outputting answer, code and suggested related function names. The answer is already a natural language response and is displayed to the student. Code is then processed by another LLM function to generate pseudo-code before being displayed to the student as interactive pseudo-code. Lastly, relevant function names go through a key/value, look-up table and retrieve any matching function documentation.}
    \caption{Revised prompt design and system architecture illustrating the data flow of the \textit{General Question} feature, highlighting the process of generating pseudo-code and relevant function documentation.}
    \label{fig:updated_prompt_architecture}
\end{figure*}

We redesigned the feature based on feedback so that it would highlight the specific lines that require modifications, deletions, or additions. To achieve this, we used a data flow which is displayed in Figure \ref{fig:updated_prompt_architecture}. (1) \textbf{Pre-processing: }The buggy code is stripped of any comments and is then reformatted with a standard style. (2) \textbf{Generating Fixed Code: }An LLM function tries to generate the fixed version of the buggy code based on the provided intended behaviour or error message. This step also generates a paragraph of changes similar to the initial version of the feature which is immediately streamed to the client and displayed. (3) \textbf{Matching Lines: }A simple static code analyzer matches each line of the fixed code to the original buggy code. (4) \textbf{Annotating Buggy Code: }The buggy code is then annotated with three labels: \raisebox{-2.5pt}{\includegraphics[scale=0.7]{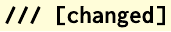}}, \raisebox{-2.5pt}{\includegraphics[scale=0.7]{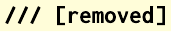}}, or a new empty line with \raisebox{-2.5pt}{\includegraphics[scale=0.7]{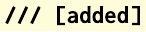}}. (5) \textbf{Explaining Annotations: }The annotated buggy code and the fixed code are sent to another LLM function that adds explanations to each of the changed, removed, and added labels. Eventually, the explained changes and annotated code is streamed to the client for rendering and displaying the highlighted lines and on-hover interactions as displayed in Figure \ref{fig:v2_system_features}a.

\subsubsection{Improved Follow-Up Questions}
We redesigned the prompts to add the capability for users to ask follow-up questions in the \textit{Question from Code}, \textit{Explain Code}, and \textit{Help Write Code} features, thus improving the steering experience. Additionally, we integrated suggested follow-up prompts (Figure \ref{fig:v2_system_features}c), inspired by the \textit{"did you mean X instead?"} suggestions commonly seen in search engines.

\begin{figure*}
    \centering
    \includegraphics[width=1\textwidth]{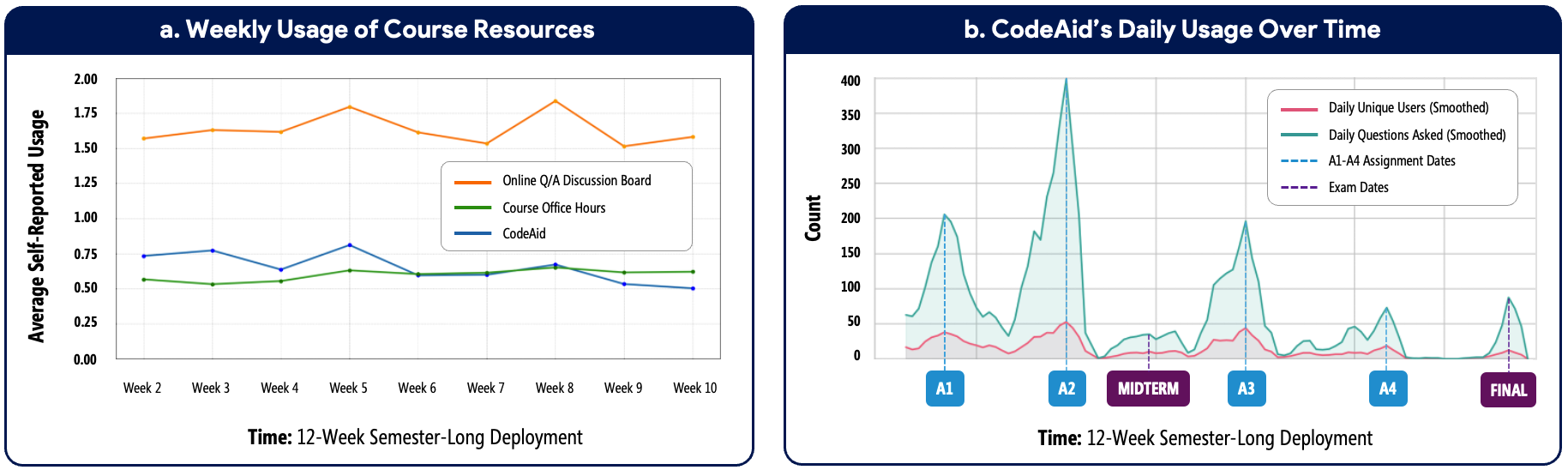}
    \Description{There are two subfigures: (a) on the left, the daily usage of CodeAid by students throughout the semester-long deployment is displayed. A red line represents the smoothed daily count of unique student users, which remains below 50 throughout the term, showing minor peaks at every assignment deadline. Conversely, a green line indicates the smoothed number of daily questions posed, seeing sharp rises on each assignment's due date, with the most significant surge during assignment 2, nearing 350 distinct questions in a day. Additionally, the chart emphasizes the deadlines for four assignments, the midterm exam date, and the date for the final examination. and (b) on the right, the average self-reported usage of various resources by students over a 12-week semester is illustrated. An orange line depicting Q\&A Discussion Boards shows variable usage, averaging around a score of 2. Meanwhile, a blue line for CodeAid and a green line for office hours both hover around an average rating of approximately 0.75.}
    \caption{\textbf{a)} Students' self-reported average usage of various course resources over the course (based on the weekly surveys), and \textbf{b)} CodeAid's daily usage and unique user counts over the course of the 12-week semester highlight a spike in activity during deadline periods (A 7-day average was applied to emphasize general trends).}
    \label{fig:codeaid_usage_over_time}
\end{figure*}

\section{Results}~\label{section6_results}
In this section, we answer the first three research questions based on the overall usage of CodeAid throughout the entire semester. Specifically, \textit{\textbf{RQ1:} students usage patterns of CodeAid}, \textit{\textbf{RQ2:} effectiveness of CodeAid in generating correct and helpful responses without revealing solutions}, and \textit{\textbf{RQ3:} students’ perceptions of CodeAid}. We synthesize results from various data sources: CodeAid interactions, thematic analyses, interviews with 22 students (S1 - S22), and both weekly and end-of-semester anonymous surveys. Where quotations came from student interviews we have indicated the subject number (e.g. S\textit{x}). Unattributed quotations came from open-ended questions on the weekly surveys. 

\subsection{RQ1: Students' Usage Patterns and Feedback}~\label{sec:results:usage_patterns_feedback}
During the course, 372 students engaged with CodeAid, posing 8132 original and 1986 follow-up queries. From these, 300 students agreed that their data could be used, which left us with 7003 original queries and responses for analysis. Students, on average, inquired CodeAid 23.3 times (SD=41.2), spanning a range from 1 to a maximum of 333 questions per student. The majority of students (n=159) asked between 1 and 9 questions, followed by 80 students who asked 10 to 29 questions, 28 students who asked 30 to 49 questions, and 34 students who asked more than 50 questions. Student engagement surged approaching major assignment deadlines or exams (Figure \ref{fig:codeaid_usage_over_time}b). The dataset of queries from consented students is available as supplementary material with this paper, offering a valuable resource for those interested in advancing AI educational tools.

We analyzed the usage of CodeAid among different gender groups, particularly the predominant self-reported gender categories. On average, those identifying as women used CodeAid 33.8 times, which was significantly more than those identifying as men, with an average usage of 18.4 times (\textit{p}=.004, \textit{d}=.34, using an independent samples t-test). This suggests that despite being underrepresented in the course (making up only 30\%), women tended to use CodeAid more frequently than men.

In terms of features, the \textit{General Question} feature was used most frequently, accounting for 38\% (n=2682) of total usages, with 683 instances leading to 1412 follow-up questions. The \textit{Question from Code} feature followed at 28\% (n=1959), leading to 187 follow-ups from 108 usages. The \textit{Help Fix Code} feature was used 1611 times (23\%). The \textit{Explain Code} accounted for 5.5\% (n=388) of usages and resulted in 34 follow-up questions. Lastly, the \textit{Help Write Code} feature represented 4\% (n=283) of usages, with 36 instances prompting 48 follow-up questions. Refer to Table \ref{tab:result_usage_stats} for a summary of descriptive statistics that compare the usage count and usefulness rating of features before and after the midterm update. Additionally, based on the weekly surveys, students reported similar weekly average usages of CodeAid and course office hours (Figure \ref{fig:codeaid_usage_over_time}a).

In order to further explore students' usage patterns with CodeAid, we turn to our thematic analysis of the 1749 usages which revealed four major types of inquiries: (i) \textit{asking programming questions}, (ii) \textit{debugging code}, (iii) \textit{writing code}, and (iv) \textit{explaining code}. Students used a combination of different features to perform the above inquiries. In the sections below we report the frequency, nature of questions, and students' choices of features.

\subsubsection{Asking Programming Questions} 
The most frequent type of inquiry was programming questions with 36\% (n=643) instances from all thematically analyzed data. We classified student questions into three categories: \textit{code and conceptual clarification} 26\% (n=453), \textit{function-specification queries} 5\% (n=96), and \textit{code execution probes} 5\% (n=94). For \textit{code and conceptual clarification}, students mostly used the \textit{General Question} feature (n=323), inquiring conceptual information about syntax, pointers, string operations, data structures, input/output operations, system calls, shell programming, and compilation tools. The \textit{Question from Code} feature was also used (n=124) to understand or clarify the \textit{role}, \textit{behaviour}, and \textit{details} of particular parts in their provided code snippets. For \textit{function-specific queries}, students mostly used the \textit{General Question} feature, seeking insights into a specific function's usage, behaviour, arguments, and return types. For \textit{code execution probes}, students predominantly utilized the \textit{Question from Code} feature (n=83), and occasionally the \textit{General Question} for shorter code snippets (n=10). They used CodeAid similar to a compiler to verify execution, evaluate output on particular inputs, and check for errors.

\begin{table*}[ht]
\centering
\caption{Summary of usage count and average usefulness ratings for each feature, broken down by version (V1: pre-midterm update, V2: post-midterm update)}
\begin{tabular}{lccccc}
\hline
Feature Type & Count V1 & Rating V1 & Count V2 & Rating V2 \\
\hline
General Question & 1648 & $M=3.99$, $SD=1.34$ & 1034 & $M=4.10$, $SD=1.32$ \\
Question from Code & 1526 & $M=3.28$, $SD=1.57$ & 433 & $M=3.27$, $SD=1.57$ \\
Help Fix Code & 1348 & $M=2.88$, $SD=1.61$ & 263 & $M=2.38$, $SD=1.51$ \\
Explain Code & 296 & $M=4.19$, $SD=1.26$ & 92 & $M=4.16$, $SD=1.15$ \\
Help Write Code & 98 & $M=3.38$, $SD=1.62$ & 185 & $M=3.25$, $SD=1.52$ \\
\hline
\end{tabular}
\Description{The table presents descriptive statistics organized by Feature Type and Version. The header lists the feature type, count for version 1 (v1), mean and standard deviation of rating for v1, count for version 2 (v2), and mean and standard deviation of rating for v2. The feature types encompass categories such as General Question, Question from Code, Help Fix Code, Explain Code, and Help Write Code.}
\label{tab:result_usage_stats}
\end{table*}

\subsubsection{Debugging Code} 
The second most frequent usage of CodeAid was to debug code (32\%). We discovered three major types of inquiries: \textit{buggy code resolution} 22\% (n=385), \textit{problem source identification} 7\% (n=130), and \textit{error message interpretation} 3\% (n=50). For \textit{buggy code resolution}, students predominantly used the \textit{Help Fix Code} (n=385) by providing their erroneous code, its intended behaviour, and sometimes, the encountered error message. Students also used \textit{Question from Code} (n=60) by providing their erroneous code and asked how to fix the error.  For \textit{problem source identification}, students used \textit{Question from Code} and specifically asked about why they were getting an error (n=108). Similarly, they used the \textit{General Question} (n=16) but for smaller contexts (e.g., one line of codes, or shell programming commands), or by mistake (e.g., forgetting to include their code by not using \textit{Question from Code}). The most common type of inquiry here was to understand functional inconsistencies such as unexpected outputs, behaviours, warnings, and specific error messages. Students were seeking explanations for why their code behaved differently than expected (e.g. \textit{"why does my truncate function not change str?"}). Lastly, for \textit{error message interpretation} students used \textit{Question from Code} (n=34) and \textit{General Question} (n=11) to understand syntax errors, Valgrind's memory-related error summaries, and command-line errors.

\subsubsection{Writing Code} 
The third most common usage of CodeAid was assisting with writing code (24\%). Students were asking for \textit{high-level coding guidance} 13\% (n=237), or \textit{direct code solutions} 10\% (n=185). In the context of seeking \textit{high-level coding guidance}, our thematic analysis reveals that students predominantly utilized the \textit{General Question} feature (n=206), followed by \textit{Question from Code} (n=19) and \textit{Help Write Code} (n=13). Typical inquiries were characterized by "how-to" questions, where students sought information on appropriate functions for specific tasks, such as \textit{"how to get the length of a string?"}, or step-by-step guidance on specific behaviours, like \textit{"How do you check if a file exists?"}. When asking for \textit{direct code solutions}, students used \textit{General Question} (n=73), followed by \textit{Help Write Code}, and \textit{Question from Code} (n=39). 

\subsubsection{Explaining Code} 
Lastly, students used CodeAid to explain the starter code that was provided to them by the course staff 6\% (n=97) using the \textit{Explain Code} feature.

\subsection{RQ2: CodeAid's Response Quality}~\label{sec:result:per_feature_evaluation}
This section reports results from the thematic analysis regarding correctness, helpfulness, and the extent to which responses directly revealed code solutions. Additionally, we report student ratings regarding usefulness and their reasons corresponding to high and low ratings.

\subsubsection{Overall Correctness and Helpfulness}
Based on our thematic analysis of 1,749 random samples, we found that the correctness rate was 79\% (1,386 correct instances) and the helpfulness rate was 86\% (1,196 out of 1,386). Notably, after updating CodeAid, there was an improvement in the quality of responses. Correctness of responses increased from 74\% (781 out of 1,057) to 87\% (603 out of 692), and their helpfulness rose from 83\% (646 out of 781) to 91\% (550 out of 603).

\subsubsection{Not Displaying Direct Code Solutions}
The assistant succeeded in avoiding the display of direct code solutions. In response to 43\% of queries, CodeAid produced purely natural language answers which included conceptual explanations. For 24\% of queries, it produced pseudo-code, of which 16\% were high-level and generic example codes, while 6\% were the pseudo-code that implemented a specific behavior. These specific pseudo-codes, although indirect, might have revealed the high-level ideas about implementing a particular behavior that was required in an assignment. Furthermore, when debugging code using CodeAid, it never displayed the fixed code and only recommended suggestions to fix minor syntax errors (16\%) and semantic issues (8\%). Similarly, the \textit{Help Fix Code} never generated any code. However, the initial version of \textit{General Question} and \textit{Question from Code} produced generic, high-level example code in 104 instances (6\%) which did not directly implement any part of assignments and were similar to what students can find on websites like Stack Overflow. Finally, in 37 instances (2\%) these two features generated a short code solution (1-3 lines) to a specific behavior.

\subsubsection{General Question}
Our thematic analysis of 733 randomly sampled usages of the \textit{General Question} feature revealed 91\% (n=668) correct and 84\% (n=613) helpful responses. Based on student ratings, the feature's usefulness was rated highly at 4.04 (SD=1.30) on a 5-point scale. Reasons associated with highly rated responses tended to report that the response was \textit{"correct"}, \textit{"helpful"}, \textit{"concise"}, and \textit{"clear"}, as evidenced by comments like \textit{"explained my misunderstanding perfectly"}. Conversely, when giving negative ratings, students tended to report that they were due to (i) incomplete or superficial explanations, (ii) the absence of example code, (iii) irrelevant or unclear responses, or (iv) incorrect or misleading information especially with more complex requests.

\subsubsection{Question from Code}
Our thematic analysis of 467 randomly sampled \textit{Question from Code} usages revealed a slightly lower accuracy, with 66\% (n=310) being correct and 55\% (n=258) helpful. The average rating for the usefulness of this feature's responses was 3.28 (\textit{SD}=1.57). Comments that correlated with highly-rated usages included reasons such as: (i) precisely identifying and locating errors in code, exemplified by feedback like, \textit{"I had been staring at the code for so long. sometimes you just need an extra set of eyes."}, (ii) providing detailed and accurate answers, and (iii) confirming that the code compiles correctly. Student comments associated with lower ratings mentioned reasons like (i) being incorrect, incomplete, or suggesting redundant code changes, (ii) vaguely and poorly explaining responses, and (iii) CodeAid's inability to understand the code the student had provided.

\subsubsection{Help Fix Code}
Our thematic analysis of 340 instances of \textit{Help Fix Code} indicated 63\% (n=214) correct responses, with 42\% (n=142) that were deemed helpful. The average rating of this feature was lower than other features at 2.67 (SD=1.55). Looking at the in-situ feedback, when students rated this feature highly, they mentioned reasons such as providing helpful fix suggestions, correctly explaining errors (e.g., \textit{"it did a good job of explaining what was wrong"}), or confirming the absence of errors and suggesting external issues. However, the feature was occasionally deemed \textit{not} useful. Lower ratings were often associated with feedback that pointed out inaccurate or irrelevant suggestions or misinterpretations of the code's intent. Several students reported in the weekly surveys and interviews that they favoured the updated visual annotations for this feature and mentioned \textit{"now I can see where to fix the code,"} and \textit{"it highlights areas I can fix in red, which is visually very helpful".} However, they also reported challenges for more complex coding tasks introduced at the end of the course. One participant (S21) reported CodeAid became \textit{"more difficult to use for longer codes"} and could not understand the \textit{"interactions between multiple files."} 

\subsubsection{Explain Code}
From the thematic analysis of 95 randomly sampled \textit{Explain Code} usages, we discovered that 95\% (90 out of 95) of the explanations were accurate and perceived beneficial when they were correct. In terms of usefulness rating, the feature was well-received by students, obtaining an average rating of 4.17 (\textit{SD}=1.21). Highly rated responses of this feature were linked to reasons such as being \textit{"accurate"}, or providing \textit{"a useful breakdown of the code"} as well as helping them in code review and double-checking their code. Reasons that correspond to lower ratings mentioned explanations being \textit{"inaccurate"}, \textit{"too short"}, or that it \textit{"did not tell anything new about the code"}

\subsubsection{Help Write Code}
From the thematic analysis of 77 instances of \textit{Help Write Code}, the feature predominantly produced correct (92\%) and helpful (82\%) responses. In 53\% of cases, CodeAid generated the exact solution of a requested behavior and in 32\% (n=25) it generated pseudo-code for a high-level example. The feature received an average usefulness rating of 3.29 (\textit{SD}=1.56). Several students that rated the feature highly, tended to report that the feature was useful in initiating coding tasks by breaking it into smaller bits. Responses that were poorly rated included reasons such as generating incomplete or irrelevant answers.

\subsection{RQ3: Students Perspectives and Concerns}~\label{sec:result:students_perspectives_concerns}
This section presents students' experiences with CodeAid across its spectrum, and broader perspectives on AI programming assistants. 

\subsubsection{Accessibility and Convenience}
Many students appreciated CodeAid's 24/7 availability, with one noting, \textit{"I like that it's always there if I need any help."} They highlighted the private space it offers where they can \textit{"ask a lot of questions"} without \textit{"having to talk to a human who will judge"} them. Students highlighted CodeAid's role as a crucial supplementary resource to assist with coding tasks, with a student commenting, \textit{"It helped me solve issues with my code that I wouldn’t have been able to figure out on my own."}

\subsubsection{Contextual Assistance}
Students appreciated CodeAid's ability to provide \textit{"faster access to relevant knowledge"} by offering contextually relevant assistance and \textit{"specific solutions"} that are \textit{"more concise."} They compared CodeAid to search engines where they \textit{"can't paste code into,"} and mentioned \textit{"I like that I can word questions how I think about them rather than thinking about what the header of the most relevant stack overflow post will be."} Students found CodeAid's responses tailored to the context of the course requirements, with S10 highlighting that \textit{"CodeAid was more related to our course, ChatGPT sometimes used functions that were not used in the course."} Negative experiences included when \textit{"the AI did not understand what [they] asked"} in which they had to search online (S12, S13), or the limits placed on input length.

\subsubsection{Learning and Dependency}
Some students expressed that CodeAid has deepened their understanding, noting it \textit{"explains things more deeply for someone who is trying to learn"} (S9) and offers \textit{"a new way to learn code."} However, some students preferred indirect responses to enable deeper engagement: \textit{"I would like a hint rather than the answer."} Although some students like S14 felt that they \textit{"over-relied on it too much rather than thinking"}, many students displayed signs of self-regulation. They \textit{"never tried to get the system to show the solution"} (S2), \textit{"did not use the fix code feature"} (S4), or \textit{"ask[ed] too general questions"} (S7) so they could learn.

\subsubsection{Trust and Reliability}
Many students recognized CodeAid's utility,  comparing its accuracy to Teaching Assistants: \textit{"80\% accurate answers, similar to TA."} Students acknowledged the utility of CodeAid's assistance in specific contexts, finding it more accurate on simple questions. Some pointed out CodeAid's confident tone when producing wrong answers, as noted by a student: \textit{"it can lie to you, and still sound confident."} In terms of trust, some found it superior, noting it seemed \textit{"like a person who knows everything,"} while others expressed that they \textit{"don't trust a computer to give [them] accurate responses."} Of interest, S13 noted that they trusted CodeAid more than Google, \textit{"just because it was part of the course"} and endorsed by the instructor.

\subsubsection{Reasons for Not Using CodeAid}
Several themes emerged when students explained instances when they did not use CodeAid. The primary reason was a perceived \textit{"lack of need,"} as many found existing course materials \textit{"sufficient"} or the coursework easy. Some were either unaware of CodeAid or forgot to use it. Students also cited a preference for existing resources like GDB debugger, Stack Overflow, and Q\&A discussion boards. A few favoured ChatGPT's GPT-4 version for its ease and versatility. Personal desire for self-reliance was also a reason, with statements like \textit{"I enjoy finding solutions by myself"}. Skepticism towards AI-generated content and past negative experiences also reduced trust in the tool. Some students preferred to consult friends, and several students mentioned the mandatory feedback as a reason for not using CodeAid.

\subsubsection{Comparing CodeAid with ChatGPT}
We conducted a fully anonymous post-course survey to ask students about their usage of ChatGPT during the course, despite the course policy to avoid its use. Of the 200 respondents, 23\% exclusively used CodeAid, 38\% used both CodeAid and ChatGPT, and 19\% only used ChatGPT. The number of students who used ChatGPT \textit{"sometimes,"} \textit{"often,"} or \textit{"a lot"} was 90, compared to 66 students for CodeAid. Among the reasons for not using ChatGPT were satisfaction with existing resources, concerns over academic integrity, and doubts about its reliability. Comparatively, students appreciated ChatGPT's user-friendly interface, greater character limit, and free-form editing. They found it useful for handling complex inputs with \textit{"multiple parts"} and \textit{"using any format"}. ChatGPT was favoured for offering in-depth code reviews, generating more comprehensive answers, and producing code examples. When it came to learning about C programming concepts, students who used both tools reported a higher learning experience with ChatGPT. However, some students that used ChatGPT expressed that the direct solutions generated by ChatGPT was \textit{"not good for learning"} (S10) or that "ChatGPT sometimes does a bit too much". Similarly, S21, \textit{"I don't think that I'm learning as much as spending time to fix [the code] myself."} Conversely, students felt that they learn more using CodeAid \textit{"since it's targeted towards CS students and explanations are more technical and they do make you think"}.

\subsubsection{Future Integration of AI Coding Tools}
Most students reported that they will continue using AI coding tools, expressing that AI helps them \textit{"work more efficiently,"} and understand coding concepts in a summarized way. Several envisioned using AI to \textit{"create the skeleton code"} of their projects, \textit{"optimize [their] code,"} and handle the \textit{"tedious programming tasks that are not too complicated"} for them. Others wished to have these tools integrated into their coding environment. One student was eager to take \textit{"a class about writing prompts to get more accurate answers."} However, some students did not want to integrate AI coding tools in the near future. One mentioned limitations such as not being effective in debugging \textit{"without seeing the entire program,"} and another student mentioned that AI should not be used for learning due to its \textit{"confident but incorrect answers"} and \textit{"that it does not encourage learning."} Lastly, a student highlighted the essence of learning as \textit{"figuring things out on your own by googling, manually fixing bugs, looking at tutorials, etc."}

\section{Educator Interviews}
To gain further insights into how educators would use CodeAid in their programming classes, we conducted semi-structured interviews with eight computing educators (T1 - T8). Educators were from six countries, including Germany (T6), India (T1), Jordan (T3), New Zealand (T2), the Netherlands (T8), and the USA (T4, T5, T7). These educators were actively engaged in teaching undergraduate-level programming courses. Six of them (T1, T3, and T5-T8) had over ten years of teaching experience, one had 5-10 years (T2), and another had 3 years (T4). Most were also engaged in computer science education (CSEd) research. The interviews began by exploring the educators' backgrounds and their current challenges and strategies, especially around students' utilization of LLM-based coding tools like ChatGPT and Github Copilot. Subsequently, we introduced our pedagogical AI coding assistant, discussing its capabilities and insights gathered from our semester-long deployment as summarized in Section~\ref{section6_results}. The conversation then shifted to the educators' opinions on our tool: what they liked and disliked, their pedagogical and ethical considerations regarding its use, their interest and requirements for integrating it into their courses, and how they perceived it relative to tools like ChatGPT. Each interview, for which informed consent was obtained, was conducted over Zoom and lasted approximately one hour.

\subsection{RQ4: Educator's Perspectives}
Educators expressed varying degrees of concern about the impact of AI coding tools like ChatGPT on the classes they taught. While T5 didn't see ChatGPT as a significant issue for advanced Computer Science courses, there was a general agreement about its potential threat to introductory programming classes. Notably, T1 mentioned \textit{"I would encourage students to use a tool that respects that there’s a learner at the other end, not ChatGPT"}.

\subsubsection{General Impressions}
Educators generally held favourable impressions of CodeAid. T3 emphasized CodeAid's pedagogical approach, stating that it offered an \textit{"honest way of using ChatGPT,"}, particularly for students keen on academic integrity. Similarly, T4 mentioned that CodeAid was \textit{"the most sensible path"} and a safer alternative to \textit{"a completely unsafe and unmoderated"} tool like ChatGPT. Similarly, T5 compared CodeAid to an \textit{"excellent TA"} that prompts students to think critically rather than offering direct answers. Furthermore, T2 envisioned that CodeAid can greatly assist students in a flipped classroom setting, help students arrive at class more prepared, offer moments for self-reflection on lecture material, and support them in tackling assignments.

\subsubsection{Perceptions on Pseudo-Code Usage}
Most educators appreciated the design of the pseudo-code feature, especially the line-by-line on-hover explanations. T6 liked the way it \textit{"provides structure"} without \textit{"giving away the difficulty of the syntax."} Similarly, T2 and T4 mentioned how it reduces cognitive load by focusing on overall logic. T4 expressed that \textit{"hiding the syntax"} helps with students' meta-cognitive skills. From an ethical standpoint, T1 claimed that showing pseudo-code was even \textit{"better than Google"} for certain queries, as opposed to viewing \textit{"precise solutions available on Stack Overflow"}. However, T3 expressed slight concerns, particularly for upper-level courses, where revealing the algorithm via pseudo-code would be detrimental.

\subsubsection{Concerns about Incorrect Responses and Misuses}
Despite positive impressions, educators expressed various concerns. After viewing the results from our thematic analysis, T1 pointed out the risk of incorrect responses, especially for students whose fundamentals are poor. T3 raised concerns about students \textit{"trusting whatever the AI says"}. T1 suggested that these tools should \textit{"build up students' ability to critique"} LLM-generated responses. As a solution, T1 and T3 suggested including mandatory tutorials with quizzes before students can use CodeAid. T5 asked for more transparency by having CodeAid display recent incorrect responses for each of the features. T3 mentioned that the inaccuracies of the \textit{Help Fix Code} feature could mislead students by highlighting incorrect lines, and suggested the feature could instead highlight potentially incorrect lines and ask self-reflective questions from students like \textit{"Are you sure this line is doing [X]?"}. Additionally, T1, T4, T6, and T7 voiced worries about misusing or abusing CodeAid through repetitive queries and suggested throttling usage as a potential solution.

\subsubsection{Keeping Students from Switching to ChatGPT}
Many educators felt that CodeAid should be designed in a way to keep students from switching to ChatGPT which \textit{"was just a click away"}. T7 mentioned that \textit{"I can't prevent students from using ChatGPT, but if I can get more students to use this tool instead of ChatGPT, then that's better"}. As a way to attract students, T2 suggested \textit{"creating an all-encompassing tool"}. Similarly, T7 and T8 suggested including a complete code editor with code execution capabilities to make it easier for students to remain engaged with CodeAid rather than defaulting to ChatGPT. Another recurring suggestion was revealing code solutions after multiple failed attempts to prevent frustration. For example, T1, T2, and T5 proposed gamifying the experience such as \textit{"showing code could cost them some kind of points in the system"} or T7 mentioned to "lock the system" after showing code, asking the student to do something useful like explaining the answer. However, T3 strongly favoured CodeAid's pedagogical approach and did not want CodeAid to reveal code solutions, asserting that \textit{"if a TA is controlled to answer in an appropriate way and not show the solution, then this tool should also be controlled"}. When discussing the ability to customize CodeAid's responses during a course, T4 was confronted with the dilemma that if they turned off a feature, then students might default to ChatGPT.

\subsubsection{Pedagogical Customization}
A recurrent theme centred on the importance of customization by instructors. Both T3 and T4 emphasized the need for instructors to have control over the types of responses generated by CodeAid for different types of questions. Specifically, T3 wanted CodeAid to produce pseudo-code only for implementation questions, and to merely offer hints for problem-solving questions. On the other hand, T4 wanted to control when pseudo-code was displayed and only enable it at the beginning of the course. T7 wanted to update CodeAid with a list of topics that have been taught in class so that it would not use other complex topics and functions when responding to students' queries.

\subsubsection{Student Monitoring Dashboard}
Another prevalent theme was the need for instructor dashboards that monitor student interactions and track their queries. T2 highlighted that by tracking students, we can see what type of questions they are asking, and what type of answers are being produced. This data could help educators in identifying gaps in their instruction, as evidenced by frequently asked questions. Such insights might prompt them to \textit{"step in and provide better examples"}. However, this monitoring comes with its own set of challenges. While T6 suggested aggregated data could provide feedback, accessing individual data might be restricted due to regulations such as the General Data Protection Regulation (GDPR). This poses a question on the balance between personalization and privacy. Lastly, T8 mentioned a crucial point concerning student anonymity and comfort. While recognizing the potential pedagogical benefits of understanding student queries, she mentioned \textit{"Students should not feel like someone is watching them and they should feel the liberty to ask anything"}.

\section{Beyond CodeAid: Implications for Pedagogical LLM-Powered Coding Assistants}
The iterative development of CodeAid and insights gained from its semester-long deployment enable us to propose design implications for the broader design of AI assistants in educational contexts. We position these implications within four main stages of a student's help-seeking process with an AI assistant: 1) The decision to use the AI tool; 2) The formulation of a query; 3) The nature of response that is supplied; and 4) Actions needed once a response is received. Our results point to four high-level design considerations, for each of these stages, each with unique usability and educational trade-offs:
\begin{itemize}
    \item \textbf{D1:} \textbf{Exploiting Unique Advantages of AI.} For deciding when to use the tool, what is the role and \textit{unique advantages} of an AI assistant compared to other available resources within the learning ecosystem?
    \item     \textbf{D2:} \textbf{Designing the AI Querying Interface}. What are the UI considerations for an AI assistant to allow users to formulate queries/prompts in a way that balances user-friendliness with meta-cognitive engagement?
    \item \textbf{D3:} \textbf{Balancing the Directness of AI Responses.} How direct should the AI assistant's responses be, so that it balances directness and learning engagement, and who should control this balance?
    \item \textbf{D4:} \textbf{Supporting Trust, Transparency and Control.} Once a response from an AI assistant is received, what UI considerations are needed to ensure accuracy, trust, transparency, and control?
\end{itemize}

\begin{figure*}
   \centering
    \includegraphics[width=1\textwidth]{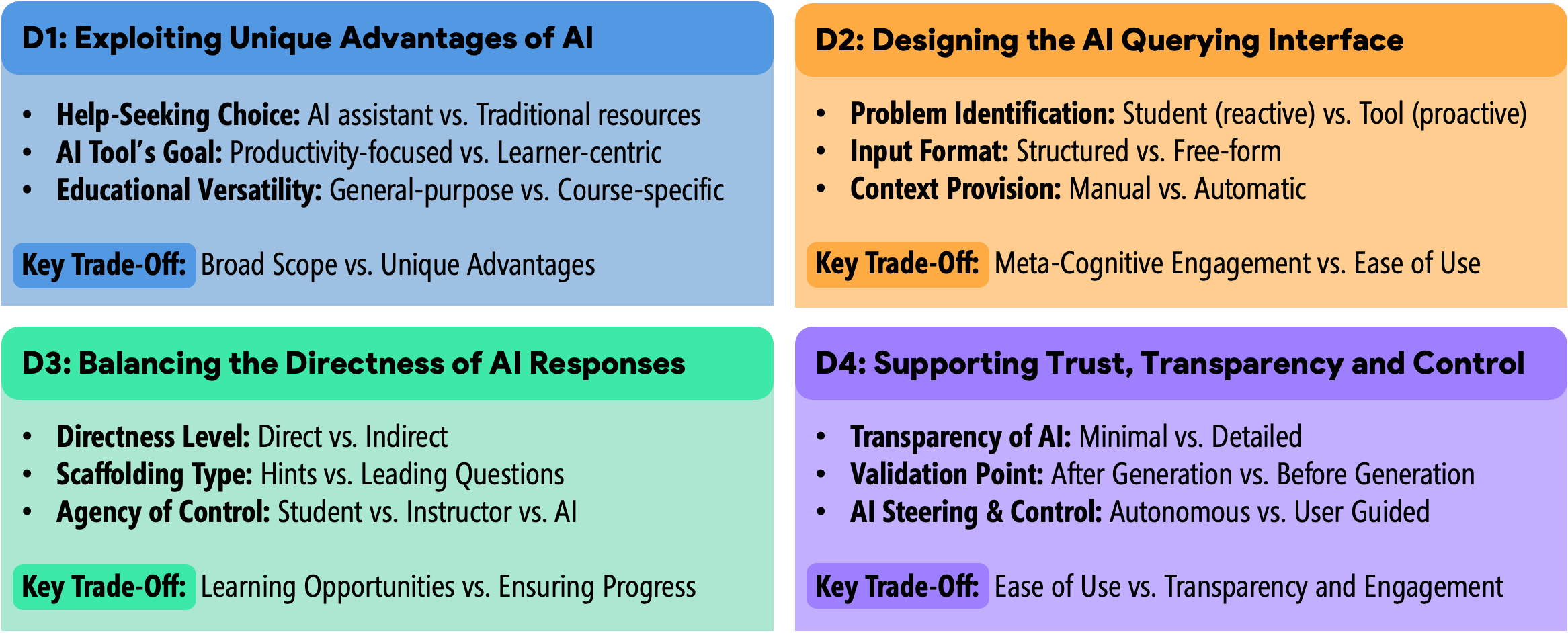}
    \Description{This figure contains four sections discussing design considerations for AI educational tools. Section D1, in blue, covers the unique advantages of AI, such as help-seeking choices and the balance between broad AI capabilities and specific benefits. Section D2, in orange, addresses the AI querying interface, focusing on problem identification, context gathering, and balancing user-friendliness with cognitive engagement. Section D3, in green, debates the directness of AI responses and the balance between scaffolding and learner autonomy. Finally, section D4, in purple, delves into trust, transparency, and control post-response, weighing transparency against ease of use. Each section includes key trade-offs highlighting essential decisions in the design process.}
    \caption{Emerging design considerations and trade-offs within the design space of AI-powered assistants for educational settings. Each consideration is based on a key stage in students' help-seeking process.}
    \label{fig:design_dimensions}
    
\end{figure*}

\subsection{D1: Exploiting Unique Advantages of AI}
An initial stage of the help-seeking process is deciding what learning resource to use within the learning material ecosystem that may be available to them. This leads to the first major design consideration of an AI coding assistant (Figure \ref{fig:design_dimensions}, D1): determining the role, scope, and unique advantages of future educational AI assistants in relation to other educational resources, like TA office hours, discussion boards, textbooks, etc. In our study, students appreciated several features unique to AI assistants, such as the ability to interact with the tool in natural language, the tool's ability to provide contextual assistance, and tailored responses. More specific to CodeAid, students pinpointed unique advantages such as its stimulation of critical thinking (using pseudo-code), technical focus on C programming (relevant to their course), and its capability to "pull" comprehensive documentation related to queries. Similar features should be built within future tools. 

Considering the rapidly expanding ecosystem of productivity-focused AI tools, it is challenging to keep students in a learner-centric AI tool. To address this issue, our results suggest designing course-specific AI assistants, for example using approaches such as retrieval augmented generation (RAG) to allow specific contextual information to be included in model responses \cite{lewis2020retrieval}. These assistants should be capable of (a) generating accurate, technically correct, and informative responses, including references to specific lecture notes, and (b) allowing students to ask detailed questions about their course projects, assignments, and logistics. Furthermore, with access to the curriculum, the AI assistants' responses will align more closely with students' zone of proximal development \cite{vygotsky1978mind}, making them more relevant and beneficial for the students. Future tools can also direct students to content on the web such as relevant StackOverflow posts, similar to Bing Chat, and relevant documentation as illustrated in CodeAid, as well as providing access to instructor-verified practice exercises and sub-goal-labeled worked examples \cite{margulieux2012subgoal}.

\subsection{D2: Designing the AI Querying Interface}
Once a user has chosen to use an AI coding assistant, the next main stage will be to enter their query. In CodeAid this is done through a structured user interface, whereas in ChatGPT, users enter free-form input. In any future system, another major design consideration is how the querying interface should be designed, such that it balances user-friendliness with meta-cognitive engagement (Figure \ref{fig:design_dimensions}, D2). Some students in our study preferred the simplicity and free-form nature of ChatGPT, while educators appreciated CodeAid's structured design, and found that it promoted active engagement, critical thinking, and thoughtful inquiry.

For instance, designers should decide (a) whether the assistant automatically gathers context (like a plugin in an IDE) or requires manual input from the user, (b) whether the assistant allows students to ask questions in a free-form manner, similar to interaction with popular chat-based AI models, or requires input in a specific structure and format, and (c) the assistant's level of reactivity versus proactivity in identifying problems and facilitating help-seeking. The trade-off to consider is that automatic context integration, free-form input methods, and proactive assistance may enhance usability and ease of use, but could potentially reduce opportunities for students to engage in meta-cognitive activities -- such as self-reflection and asking well-formed questions that promote critical thinking -- which are essential for success in learning programming \cite{prather2020what}. Future research is also needed to evaluate the impact of proactive problem identification and assistance, focusing on the design of minimally distracting assistants. Indeed, recent work by Prather et al. revealed that students did not like being shown suggestions when they felt they did not need the help \cite{prather2023weird}.

\subsection{D3: Balancing the Directness of AI Responses}
After formulating the question, the degree of control over the type and directness of the assistance is the next design consideration (Figure \ref{fig:design_dimensions}, D3). According to our study, students and educators had diverse views and requirements regarding the level of directness and scaffolding an AI assistant should offer. Students in our study exhibited varying needs: some sought explicit example code, others preferred subtle hints, and there were instances where they simply wanted the AI to provide direct solutions. Furthermore, educators highlighted the importance of customization options, allowing them to tailor the AI's level of assistance, such as restricting the display of pseudo-code for specific queries or during certain course segments, to better align with pedagogical goals. This introduces an important trade-off: finding the right balance between sufficient scaffolding for critical learning and minimizing frustration, while also considering the degree of autonomy learners should have in personalizing their level of scaffolding.

If the assistant is too direct and students have high autonomy over selecting their desired level of scaffolding, students might miss out on critical learning opportunities tailored to their zone of proximal development. This might negatively impact skill development and self-efficacy. Conversely, indirect responses can risk discouraging students if they feel overwhelmed, unsupported, and not making progress.

To promote deeper cognitive engagement and ensure progress, future tools might, as a last resort, display generated code but lock the AI tool until students complete a specific learning activity. This could involve the AI assistant highlighting critical sections of the code and prompting students to answer comprehension questions about these parts. Additionally, future AI tools could foster independent problem-solving using the \textit{Socratic Method} \cite{tamang2021comparative} and transforming code generation requests into a series of problem-solving questions. With each correct response, the tool would reveal the code segment corresponding to that question. 

Furthermore, since code examples are a crucial learning resource \cite{loksa2016role}, future AI assistants can differentiate between example code and direct solutions. Our thematic analysis revealed that a majority of queries were asking for code and conceptual clarifications, function-specification queries, error message interpretation, and high-level coding guidance. Consequently, future tools could be designed to display example code in response to queries not directly related to course assessments.

\subsection{D4: Supporting Trust, Transparency and Control}
Upon receiving a response from the AI assistant, users must evaluate its accuracy and helpfulness, and if necessary, provide more information to steer the AI towards a more suitable answer. Recent studies highlight the challenges in user interactions with generative AI models due to their low transparency and predictability \cite{amershi2019guidelines}. However, these challenges are compounded in educational AI assistants, which, like CodeAid, may be designed to provide scaffolded, indirect responses without directly revealing solutions. Our thematic analysis showed how CodeAid's responses were sometimes incorrect or unhelpful. As such, students seeking to verify CodeAid's responses often had to write and test a small program based on the provided suggestions. In contrast, students using a tool that provides direct responses, such as ChatGPT, could immediately access and run the generated code for verification. Any unpredictability within a learning context could erode trust in an AI coding assistant.

This points to a fourth design consideration (Figure \ref{fig:design_dimensions}, D4): given the indirect nature of responses in educational AI assistants, how can we maximize user experience, efficiency, and predictability of obtaining a helpful response. This could involve displaying the AI's assumptions on input queries, engaging students in verifying them, and enabling feedback mechanisms to ensure high-quality responses. While such techniques could lead to higher accuracy, they may complicate user interaction and potentially overwhelm students. The best method for validating responses after generation is an open question, and requires targeted future work. 

One potential approach for improving accuracy and user trust is the addition of a \textit{Sufficiency Check} step similar to CodeHelp \cite{liffiton2023codehelp} before generating the final response. This sufficiency check would actively engage users in refining the AI's understanding by prompting users about missing context or clarifying uncertainties. Moreover, future tools could enhance the validation of scaffolded responses by including code execution capabilities and enabling users to interact with and test the underlying code generated by the AI, without directly revealing it. This interaction could be facilitated through a line-by-line execution interface, similar to PythonTutor \cite{guo2013online}, and can use a black-box approach or represent a pseudo-code version.

\section{Limitations and Future Work}
Our findings from the deployment of CodeAid are contextualized within a second-year C programming course at a single university, which may not directly generalize to other courses and contexts. Furthermore, the perceived accuracy and utility of CodeAid, which in turn influenced students' trust and engagement, were heavily tied to the OpenAI models we used. The initial release of CodeAid with the \raisebox{-2.5pt}{\includegraphics[scale=0.7]{sections/figures/inline-images/t8-llm-code-davinci.pdf}} model from 2021 could have adversely shaped students' perceptions, particularly when compared to the subsequently adopted \raisebox{-2.5pt}{\includegraphics[scale=0.7]{sections/figures/inline-images/t9-llm-gpt-3.5.pdf}} model from 2023. This factor might account for some of CodeAid's inaccurate responses and the decline in student usage we observed over time. Additionally, the performance of these models varies between different programming languages \cite{xu2022systematic}, or even different code-related tasks (e.g. code explanation, fixing, or generation). For future work, we plan to run longitudinal studies to investigate how using pedagogical AI coding assistants affects long-term learning outcomes, competency, self-regulation abilities, and frustration levels in educational environments. More controlled studies could also be performed to directly compare these learning outcomes to the use of unrestricted LLMs like ChatGPT. 

Moreover, our results indicate that women used CodeAid more often than men. Future research should investigate additional demographic factors and examine how this finding connects to existing research on gender effects in the use of resources like TA office hours and online discussion Q\&A forums \cite{doebling2021patterns}.

Finally, the design considerations for AI assistants in educational contexts, as discussed in this paper, require further exploration. Considering the variability of educational contexts and the evolving nature of AI technologies, additional dimensions and trade-offs might emerge.

\section{Conclusion}
This paper presents the iterative design of CodeAid, an AI-powered coding assistant equipped with guardrails to prevent it from generating direct solutions in response to student queries. Instead, CodeAid provides scaffoldings, such as interactive pseudo-code, to foster engaging learning experiences. To understand how students utilize an AI-powered tutor and the broader implications of AI in scaling instructional expertise, we conducted a semester-long deployment of CodeAid in a programming class with 700 students. Our data collection included: (i) approximately 8,000 usages, coupled with students' feedback on the responses' usefulness; (ii) a thematic analysis of 1,749 usages in terms of correctness, helpfulness, and the extent of revealing direct solutions; (iii) weekly surveys and 22 semi-structured interviews with students; and (iv) a final anonymous survey focusing on the use of ChatGPT. Additionally, we interviewed eight programming educators to gather insights on the future of AI-powered educational tools. By synthesizing results from these diverse sources, we identified four high-level design considerations, including key trade-offs, in the emerging design space of educational AI tools. It is our hope the the results from our study, along with the broader design considerations which we have discussed, will help guide the future development of AI-powered codding assistants.


\bibliographystyle{ACM-Reference-Format}
\bibliography{references}

\newpage
\appendix
\onecolumn

\section{Thematic Analysis Codebook}~\label{appendix_codebook}
This appendix includes the final codebook used for the thematic analysis of students' usage of CodeAid. We analyzed each usage from two primary dimensions: \textit{Query}, and \textit{Response}.

\renewcommand{\thetable}{A1}

\begin{table*}[h]
\centering
\caption{CodeAid Thematic Analysis Codebook: \textbf{Query} - \textit{What are students asking from CodeAid?}}
\small
\renewcommand{\arraystretch}{1.5}
\begin{tabularx}{\textwidth}{>{\hsize=0.01\hsize}l>{\hsize=0.34\hsize}X|>{\hsize=0.65\hsize}X}
\specialrule{.1em}{.05em}{.05em} 
& \textbf{Dimensions and Codes} & \textbf{Code Description} \\ \hline
&\textbf{What is being asked?} & \textit{Focusing on the content of their usage, what are they asking for, or trying to do?} \\
1. & Error Message Interpretation & Students asking CodeAid about why their code is not working correctly, having errors, segmentation faults, etc. \\ 
2. & Problem Source Identification & Students asking CodeAid to help them identify the cause of the problem.\\ 
3. & Buggy Code Resolution & Students asking about how to resolve the error or bug within the provided code.  \\ 
4. & Explain Error Message & Students asking CodeAid to explain a provided error message. Students may also include code for more context. \\ 
5. & Code Execution Probes & Students using CodeAid as a compiler and ask for the result or potential error when the provided code is executed. \\ 
6. & Code and Conceptual Clarification & A general programming question that usually has the form of \textit{"how to do [X]?"} or \textit{"what does [X] do?"} \\ 
7. & Function Specification Queries & Students asking for more information about a particular function, its usage and examples. \\ 
8. & High-level Coding Guidance & Students asking about the process of doing something at a high level and looking for some implementation detail. For example, \textit{"How can I tokenize a dynamically allocated string?"} \\ 
9. & Direct Code Solution & Students explicitly asking for solutions for their labs or assignments (e.g., by copying part of the task description). \\ 
10. & Explain Code & Students asking CodeAid to explain their code. \\ 
\specialrule{.1em}{.05em}{.05em} 
\end{tabularx}
\Description{Table showing a section of the CodeAid thematic analysis codebook, focused on "what is being asked?". The listed codes are: Problem Source Identification, Buggy Code Resolution, Explain Error Message, Code Execution Probes, Code and Conceptual Clarification, Function Specification Queries, High-level Coding Guidance, Direct Code Solution, and Explain Code. Each code is accompanied by its respective description.}
\end{table*}

\renewcommand{\thetable}{A2}

\begin{table*}[h]
\centering
\small
\caption{CodeAid Thematic Analysis Codebook: \textbf{Response} - How much is CodeAid directly revealing the solution?}
\renewcommand{\arraystretch}{1.5}
\begin{tabularx}{\textwidth}{>{\hsize=0.01\hsize}l>{\hsize=0.34\hsize}X|>{\hsize=0.65\hsize}X}
\specialrule{.1em}{.05em}{.05em} 
& \textbf{Dimensions and Codes} & \textbf{Code Description} \\ \hline
& \textbf{How much directly revealing the solution?} & \parbox[t]{\hsize}{\textit{How much is CodeAid directly revealing the solution?}\\ \textit{(The codes are sorted from most revealing to least revealing)}} \\ 
1. & Exact Solution Code & Generated the code solution to a question. \\
2. & Exact Solution Pseudo-code & Similar to \textit{"Exact Solution Code"}, but in pseudo-code. \\ 
3. & Step to Fix Semantic Issue & Generated the steps required to fix semantic/logical problems, which usually need additional lines to achieve new functionality. \\ 
4. & Step to Fix Syntax Issue & Generated the steps required to fix minor syntax issues, usually needed to perform an inline fix. \\ 
5. & Step to Fix External Issue & Generated the steps to fix an issue that is not within the code, but about the compilation or execution. \\ 
6. & Example High-level Code & Generated a generic, high-level example for a function, or a generic implementation, i.e., how to construct a linked list, often available on Stack Overflow. This includes ALL occurrences of the \textit{Inline Code Exploration} feature. \\
7. & Example High-level Pseudo-code & Similar to \textit{"Example High level Code"}, but in pseudocode. This includes any occurrences of steps/instructions on how to complete. \\ 
8. & Conceptual Explanation & Generated a response that is completely in natural language. Provides conceptual explanation, clarifications, or the result of a code execution. \\ 
& n/a & Does not apply (reserved for \textit{"Explain Code"}). \\
\specialrule{.1em}{.05em}{.05em} 
\end{tabularx}
\Description{This table presents the 'Response' segment of the CodeAid thematic analysis codebook, focusing on the degree to which a code solution is revealed. It lists codes ranging from the most detailed, like 'Exact Solution Code', to the least detailed, such as 'Conceptual Explanation'. Each code is accompanied by its respective description.}
\end{table*}

\renewcommand{\thetable}{A3}

\begin{table*}[h]
\centering
\small
\caption{CodeAid Thematic Analysis Codebook: \textbf{Response} - \textit{How technically correct?}}
\renewcommand{\arraystretch}{1.5}
\begin{tabularx}{\textwidth}{>{\hsize=0.01\hsize}l>{\hsize=0.34\hsize}X|>{\hsize=0.65\hsize}X}
\specialrule{.1em}{.05em}{.05em} 
& \textbf{Dimensions and Codes} &
  \textbf{Code Description}  \\ \hline
 &
  \textbf{How technically correct?} & 
  \textit{Despite the question, how correct is the response from the tool?} \\ 
1. &
  Correct &
 Everything including the answer and its explanation is correct. \\ 
2. &
  Incorrect &
  Any part of the answer or explanation is incorrect. \\ \specialrule{.1em}{.05em}{.05em} 
\end{tabularx}
\Description{Table describes the 'response' part of the CodeAid thematic analysis codebook, specifically, How technically correct? Codes include Correct and incorrect. Followed by their description in that order.}
\end{table*}

\renewcommand{\thetable}{A4}

\begin{table*}[h]
\centering
\small
\caption{CodeAid Thematic Analysis Codebook: \textbf{Response} - \textit{How helpful if correct?}}
\renewcommand{\arraystretch}{1.5}
\begin{tabularx}{\textwidth}{>{\hsize=0.01\hsize}l>{\hsize=0.34\hsize}X|>{\hsize=0.65\hsize}X}
\specialrule{.1em}{.05em}{.05em} 
& \textbf{Dimensions and Codes} & \textbf{Code Description} \\ \hline
& \textbf{How helpful if correct?} & \textit{Is the response helpful to students? Does it guide them the right direction based on the provided query? Does it identify their potential issues? or is it completely misleading?} \\ 
1. & Helpful & Answer that allows the student to take one step further, even if it is not arriving at the final solution. \\ 
2. & Not Helpful & Answer that does not allow students to progress any further, is unrelated to their .\\ 
& n/a & Does not apply (reserved for \textit{"Incorrect"}). \\
\specialrule{.1em}{.05em}{.05em} 
\end{tabularx}
\Description{Table describes the response part of the CodeAid thematic analysis codebook, specifically, How helpful if correct? Codes include Helpful, not helpful and NA. Followed by their description in that order.}
\end{table*}

\end{document}